\documentclass[iop]{emulateapj}
\usepackage{natbib,xspace,color} 
\usepackage{verbatim}
\usepackage{rotating}
\usepackage{gensymb }
\usepackage{mathrsfs }  
\usepackage{amsmath}
\usepackage{amssymb }
\usepackage[backref,breaklinks,colorlinks,citecolor=blue]{hyperref}
\usepackage[all]{hypcap} 

\usepackage{url} 
\newcommand{\ang}{\mbox{\AA} }
\newcommand{\EBratio}{$\langle E(B-V)_{\mathrm{star}}\rangle/\langle E(B-V)_{\mathrm{gas}}\rangle$}

\shorttitle{Characterizing Dust Attenuation in Local Star Forming Galaxies}
\shortauthors{Battisti et al.}

\begin{document}

\title{Characterizing Dust Attenuation in Local Star Forming Galaxies: Inclination Effects and the 2175~\AA\ Feature}
\author{A. J. Battisti\altaffilmark{1,2}, 
D. Calzetti\altaffilmark{2},
R.-R. Chary\altaffilmark{3}
}

\altaffiltext{1}{Research School of Astronomy and Astrophysics, Australian National University, Canberra, ACT 2611, Australia; Andrew.Battisti@anu.edu.au}
\altaffiltext{2}{Department of Astronomy, University of Massachusetts, Amherst, MA 01003, USA}
\altaffiltext{3}{MS314-6, U.S. Planck Data Center, California Institute of Technology, 1200 East California Boulevard, Pasadena, CA 91125, USA}

\begin{abstract}
We characterize the influence that inclination has on the shape and normalization in average dust attenuation curves derived from a sample of $\sim$10,000 local star-forming galaxies. To do this we utilize aperture-matched multi-wavelength data from the \textit{Galaxy Evolution Explorer}, the Sloan Digital Sky Survey, the United Kingdom Infrared Telescope, and the Two Micron All-Sky Survey. We separate our sample into groups according to axial ratio ($b/a$) and obtain an estimate of their average total-to-selective attenuation $k(\lambda)$. The attenuation curves are found to be shallower at UV wavelengths with increasing inclination, whereas the shape at longer wavelengths remains unchanged. The highest inclination subpopulation ($b/a<0.42$) exhibits a NUV excess in its average selective attenuation, which, if interpreted as a 2175~\AA\ feature, is best fit with a bump strength of 17-26\% of the MW value. No excess is apparent in the average attenuation curve of lower inclination galaxies. The differential reddening between the stellar continuum and ionized gas is found to decrease with increasing inclination. We find that higher inclination galaxies have slightly higher values of $R_V$, although this is poorly constrained given the uncertainties. We outline possible explanations for these trends within a two component dust model (dense cloud+diffuse dust) and find that they can be naturally explained if carriers of the 2175~\AA\ feature are preferentially destroyed in star-forming regions (UV-bright regions).
\end{abstract} 

\section{Introduction}
The spectral energy distribution of galaxies (SEDs) are often heavily affected by the presence of interstellar dust that acts to obscure light from ultraviolet (UV) to near-infrared (NIR) wavelengths. Understanding the wavelength-dependent behavior of this dust attenuation\footnote{We define attenuation to be a combination of extinction, scattering of light into the line of sight by dust, and geometrical effects due to the star-dust geometry. Extinction is the absorption and scattering of light out of the line of sight by dust, which has no dependence on geometry.} is important because assumptions on the dust correction have a strong impact on properties derived from SED modeling \citep{conroy13}. The dust attenuation will depend on the geometry of the dust relative to the stellar distribution \citep[see review by][]{calzetti01}. Detailed modeling of nearby galaxies has shown that dust in spiral galaxies is found to be preferentially located within the disk region with scale-lengths and scale-heights typically larger and smaller than that of the stellar component, respectively \citep[e.g.,][]{xilouris99, bianchi07, deGeyter14}. Therefore, one expects that the inclination of a galaxy relative to an observer will influence the total dust attenuation because more dust is preferentially viewed along the line of sight in more edge-on galaxies.

Numerous optical and NIR studies have examined the influence that inclination has on the attenuation and observed galaxy properties by comparing more inclined galaxies to face-on galaxies, under the assumption that the intrinsic properties are independent of inclination \citep[e.g.,][]{disney89, giovanelli94, giovanelli95, masters03, masters10, driver07, unterborn&ryden08, maller09, yip10}. These studies find that the stellar continuum of edge-on galaxy disks suffer up to 1~mag of additional attenuation in the $B$ or $g$ band, relative to face-on galaxies. The relationship between the amount of attenuation as a function of inclination is seen to be highly non-linear, with the strongest effects occurring for galaxies with $b/a<0.4$ (most inclined), where $b/a$ is the observed ratio of the disk semi-minor and semi-major axes.

A prominent feature in local extinction curves is a broad absorption feature centered at 2175~\AA\ \citep[referred to as the 2175~\AA\ feature or bump;][]{draine03}, which is observed in many sightlines of the Milky Way \citep[MW;][]{cardelli89, fitzpatrick99}. This feature is attributed to carbonaceous grains \citep[e.g.,][]{draine03, bradley05, papoular&papoular09}, perhaps in the form of graphite or PAHs, although silicate carriers have also been proposed \citep[e.g.,][]{bradley05}. However, current models still have difficulty in accommodating the observed variations in the width of the feature while maintaining a nearly invariant central wavelength. In the MW, the strength and width of the feature can vary dramatically along sightlines toward different environments and this may reflect differences in the underlying size distribution and composition of the dust grains \citep[e.g.,][]{fitzpatrick04}. The 2175~\AA\ feature is also observed in the Large Magellanic Cloud \citep[LMC; e.g.,][]{gordon03} and the Andromeda galaxy \citep[M31; e.g.,][]{bianchi96,clayton15}, but is typically very weak or absent in extinction curves of the Small Magellanic Cloud \citep[SMC; e.g.,][]{gordon03}. In addition, there are noticeable differences in the strength of this feature in the average LMC curve and that of the LMC2 supershell (corresponding to the 30 Dor star-forming complex). Assuming comparable metallicity between these regions, this difference may be due to the destruction of the grain responsible for the feature within strongly star-forming regions \citep{gordon97, gordon03, fischera&dopita11}. 

Geometry, clumping, and the apparent optical depth can play a significant role in the observed strength of the 2175~\AA\ feature when seen in attenuation \citep[e.g.,][]{gordon97, witt&gordon00, gordon00, seon&draine16}, and therefore one expects that inclination can influence its appearance. Numerous studies have characterized the behavior of attenuation as a function of inclination theoretically through radiative transfer analysis \citep[e.g.,][]{silva98, pierini04, tuffs04, inoue05, jonsson10, chevallard13}. These models predict that, if adopting an underlying MW extinction curve, attenuation curves should become shallower with weaker 2175~\AA\ bumps at higher inclinations. The bump strength is also expected to decrease with increasing dust column density. 

Numerous observational studies have looked into the strength of the 2175~\AA\ feature in the attenuation curves of local galaxies, often with noticeably different results. Using UV spectroscopy of 39 local starburst galaxies (strongly star-forming galaxies, SFGs)  \citet{calzetti94}, find that the attenuation curve lacks the 2175~\AA\ feature. A complete lack of a feature cannot be entirely explained from geometric, clumping, or optical depth effects and therefore has been argued to result from the feature being absent or noticeably suppressed in the intrinsic extinction curve of these galaxies \citep{gordon97, gordon00, seon&draine16}. This may be due to grain destruction within strongly star-forming regions \citep{gordon97, gordon03, fischera&dopita11}. 

Others studies have utilized photometry to infer the presence of the 2175~\AA\ feature. \citet{conroy10} utilized a sample of $\sim$3400 local disk galaxies and compared the broadband photometry of galaxies of differing inclinations and inferred a bump feature with a strength 80\% that of the MW. \citet{wild11} utilized a pair-matching technique on a sample $\sim$10,000 local galaxies, matching them based on global properties and then comparing more dusty and less dusty galaxies as determined from the Balmer decrement ($F(\mathrm{H}\alpha)/F(\mathrm{H}\beta)$), to characterize attenuation from broadband photometry and found that more inclined galaxies tend to have shallower UV slopes and stronger 2175~\AA\ features. \citet{hagen17} performed a spatially resolved analysis of attenuation in the SMC, finding that the bump strength decreases with increasing total attenuation (optical depth), a result consistent with expectation from radiative transfer models \citep[e.g.,][]{pierini04, seon&draine16}.

The limited number of UV spectroscopic instruments to date, due to requiring space-based facilities, has made examining high redshift galaxies with optical facilities an alternate approach to explore this feature. \citet{noll09a} examined 78 galaxies at $1<z<2.5$ using optical spectroscopy and found that a bump feature with 28\% MW strength is present in 24\% of their sample, with the remaining 76\% of cases exhibiting a weaker bump feature with $\sim$15\% MW strength. \citet{buat11b} utilized intermediate-band data of 30 galaxies at $1<z<2$ and find that they require a bump with a strength of 35\% that of the MW. This was followed by \citet{buat12} using a much larger sample of 751 galaxies at $1<z<2$ that found the bump is securely detected in 20\% of their sample with an average strength of 48\% that of the MW. \citet{kriek&conroy13} also utilized intermediate-band data of $0.5<z<2$ galaxies and find that galaxies experiencing steeper attenuation curves require a stronger bump strength, with an average bump strength of $\sim$30\% that of the MW. \citet{seon&draine16} find that they can reproduce the trend of \citet{kriek&conroy13} through radiative transfer modeling if they adopt a MW-like extinction curve with a bump strength of $\sim$30\% of that of the MW dust. \citet{scoville15}, using a sample of 266 galaxies at $2<z<6$,  inferred an attenuation curve based on broadband photometry and the CIV absorption feature that has a noticeable bump feature. \citet{reddy15} determined the attenuation curve of 224 galaxies at $1.4<z<2.6$ by combining broadband photometry and Balmer decrement measurements and found that a 2175~\AA\ feature is not required, but suggests that a weak feature may exist. Several other studies have also suggested that a 2175~\AA\ feature is not evident in high redshift galaxies \citep[e.g.,][]{zeimann15, salmon16}. 

As can be gathered, there is no consensus regarding the importance of the 2175~\AA\ feature in galaxy attenuation curves, likely owing to large differences in the galaxy samples considered and the methodologies employed for characterizing the dust attenuation. Additionally, excluding \citet{wild11}, none of the previous observational studies have explicitly examined the entire UV-NIR behavior of the attenuation curve and the importance of the 2175~\AA\ feature as a function of galaxy inclination. 

The goal of this study is to extend on the work of our previous studies \citep{battisti16,battisti17} by quantifying the influence that inclination has on the shape and normalization on dust attenuation curves, as well as its role on the strength of the 2175~\AA\ feature, using a sample of $\sim$10,000 local SFGs. The benefit of this sample is that the amount of dust attenuation for each galaxy can be robustly inferred from the Balmer decrement, whereas most previous studies have employed SED modeling techniques that are dependent on the assumed stellar population and therefore subject to more uncertainty and potential biases.

Throughout this work we adopt a $\Lambda$-CDM concordance cosmological model, $H_0=70$~km/s/Mpc, $\Omega_M=0.3$, $\Omega_{\Lambda}=0.7$. To avoid confusion, we make explicit distinction between the color excess of the stellar continuum $E(B-V)_{\mathrm{star}}$, which traces the reddening of the bulk of the galaxy stellar population, and the color excess seen in the nebular gas emission $E(B-V)_{\mathrm{gas}}$, which traces the reddening of the ionized gas around massive stars within HII regions. We quantify the strength of the 2175~\AA\ feature (i.e., amplitude relative to the baseline extinction or attenuation curve) through the parameter $E_b$ following \citet{fitzpatrick&massa90, fitzpatrick&massa07}. For reference, the MW curve has an $E_b=3.30$ \citep{fitzpatrick99}, the average LMC and LMC2 supershell (30 Dor) curves have an $E_b=3.12$ and $1.64$, respectively \citep{gordon03}, and the SMC has a value of $E_b\lesssim0.39$ \citep{gordon03}. 

\section{Data and Measurements}
\subsection{Sample Selection}\label{data}
The parent sample of galaxies used in this study is the same as in \citet{battisti16}, and we refer the reader to that paper for a detailed description. In brief, galaxies are selected from the \textit{Galaxy Evolution Explorer} \citep[\textit{GALEX}; ][]{martin05,morrissey07} data release 6/7 catalogs of \citet{bianchi14} with FUV ($1344-1786$~\AA) and NUV ($1771-2831$~\AA) detection of $S/N>5$. These sources are cross-matched with spectroscopically observed galaxies in the Sloan Digital Sky Survey (SDSS) data release 7 \citep[DR7;][]{abazajian09}. We utilize optical emission line diagnostic to remove sources with a significant active galactic nucleus (AGN) contribution from the sample and also to determine the dust content (through the Balmer decrement). The sample is also chosen to be $z\le0.105$ to avoid the FUV band from being influenced by strong absorption features. Together these selections provide us with a sample of 9813 SFGs that act as our parent sample. This parent sample is predominantly composed of disk galaxies, as expected for SFGs. Using the NYU Value-Added Galaxy Catalog\footnote{\url{http://sdss.physics.nyu.edu/vagc/}} \citep[NYU-VAGC;][]{blanton05a}, we find that 88\% of the parent sample has a $r$-band S\'ersic index of $n\leq2.5$ ($n=2.5$ roughly separates disk-dominated from bulge-dominated galaxies, with larger numbers indicating bulge-dominated cases).

The UV-optical photometry is obtained using 4.5\arcsec\ diameter aperture photometry (constrained by the GALEX point-spread function) to remain as self-consistent as possible to the 3\arcsec\ diameter fiber used for the spectroscopy. This methodology is outlined in \citep{battisti16}. As a brief reminder, for the UV we use the NYU-VAGC $u$-band light profiles, assuming it to be a reasonable tracer of UV color, to determine appropriate aperture corrections for each individual galaxy. The amount of light within $4.5\arcsec$ in the modeled $u$-band profile is compared to the same measurement after being convolved with the \textit{GALEX} PSFs and the aperture correction is taken as the ratio of these values. For the SDSS spectroscopy ($3\arcsec$), we follow a similar methodology and determine the correction using the NYU-VAGC light profile models for the $gri$-bands (overlap in wavelength with the spectrum). We perform a chi-squared minimization to match the optical spectrum to the $4.5\arcsec$ model photometry. We find that these corrections result in good visual agreement between the UV and optical data. It is also worth stating that the methodology for determining attenuation curves used in this work (and previously) is not sensitive to the absolute scale of these aperture corrections because it is based on taking ratios of galaxy templates as a function of dust reddening (i.e., even if the aperture corrections are off by a constant factor, the derived curve would be identical).

In \citet{battisti17}, we cross-matched our sample with NIR sources in the UKIRT (United Kingdom Infrared Telescope) Infrared Deep Sky Survey \citep[UKIDSS; ][]{hewett06} and the Two Micron All-Sky Survey \citep[2MASS; ][]{skrutskie06}. The former survey is deeper than the latter but only encompasses a fraction of the SDSS footprint, thus using both provides us with good statistics for both faint (common) and bright (rare) galaxies. Combining the two samples provides NIR data for 5546 unique SFGs from the parent sample. The NIR photometry is matched to the same 4.5\arcsec\ diameter aperture \citep[see][for details]{battisti17}. As a brief reminder, we perform the NIR corrections using the annular light profile catalog (\texttt{PhotoProfile}) from the SDSS, which we convert into cumulative flux densities. We determine the flux densities at 4.5\arcsec , 5.7\arcsec , (UKIDSS catalog aperture), and 8.0\arcsec\ (2MASS catalog aperture) and normalize the UKIDSS and 2MASS photometry such that the ratio between the NIR and the SDSS $z$-band flux density at the catalog aperture is preserved for $4.5\arcsec$ (e.g., $f_J(4.5\arcsec)=f_z(4.5\arcsec)[f_J(5.7\arcsec)/f_z(5.7\arcsec)]$). For sources detected in both UKIDSS and 2MASS, we find this method to provide consistent values between the two surveys.

All measurements of galaxy properties utilized in this work are from the Max Planck Institute for Astrophysics and Johns Hopkins University (MPA/JHU) group\footnote{\url{http://www.mpa-garching.mpg.de/SDSS/DR7/}} and correspond to the 3\arcsec\ SDSS fiber which is typically centered on the nuclear region and represents a fraction of the total galaxy, unless this is specified otherwise. We adopt the uncertainty values listed in \citet{juneau14} for the emission lines, which are updated values for the DR7 dataset. The stellar masses are based on fits to the photometric data following the methodology of \citet{kauffmann03a} and \citet{salim07}. The star formation rates (SFRs) are based on the method presented in \citet{brinchmann04}. The gas phase metallicities are estimated using \citet{charlot&longhetti01} models as outlined in \citet{tremonti04}. All photometry and spectroscopy has been corrected for foreground Milky Way extinction using the \textit{GALEX} provided $E(B-V)_{\mathrm{MW}}$ with the extinction curve of \citet{fitzpatrick99}. For the \textit{GALEX} bands, we adopt the values of $k_{\mathrm{FUV}}=8.06$\footnote{$k(\lambda)\equiv A_\lambda /E(B-V)$ is the total-to-selective extinction.} and $k_{\mathrm{NUV}}=8.05$, which represent the average value of the MW extinction curve convolved with each filter on SEDs with UV slopes $-2.5 <\beta<0.5$, the typical range for our SFG sample.

\section{Methodology}\label{method}

\subsection{Characterizing Inclination}\label{method_inclination}
The primary focus of this study the quantification of the influence that inclination has on the shape of the attenuation curve. To properly examine this, it is important that galaxy inclinations are accurately determined. Nominally, the inclination angle, $\theta$, of a disk galaxy can be determined from
\begin{equation}
\cos^2(\theta)=\frac{(b/a)^2 - q_z^2}{1 - q_z^2} \,,
\end{equation}
where $b/a$ is the observed ratio of the semi-minor and semi-major axes and $q_z$ is the intrinsic ratio between the vertical and radial scale height of the disk. The value of $q_z$ will vary among disk galaxies, but is typically between $0.1<q_z<0.3$ \citep[e.g.,][]{unterborn&ryden08, padilla&strauss08, rodriguez&padilla13}. Due to the uncertainty in the vertical scale height of disk galaxies, we choose to adopt the axial ratio as our proxy for inclination. However, we will still refer to galaxies at low axial ratios as higher inclination galaxies (closer to edge-on, $\theta=90^{\circ}$) and high axial ratios as low inclination galaxies (closer to face-on, $\theta=0^{\circ}$). The ellipticity of the disk relative to the intrinsic axial ratio viewed face-on, $\epsilon=1-b_0/a_0$, can also affect the estimation of inclination, with typical values for spiral galaxies between $0.0<\epsilon<0.2$ \citep[e.g.,][]{unterborn&ryden08, padilla&strauss08, rodriguez&padilla13}.

\begin{figure*}
\begin{center}
$\begin{array}{cc}
\includegraphics[scale=0.5]{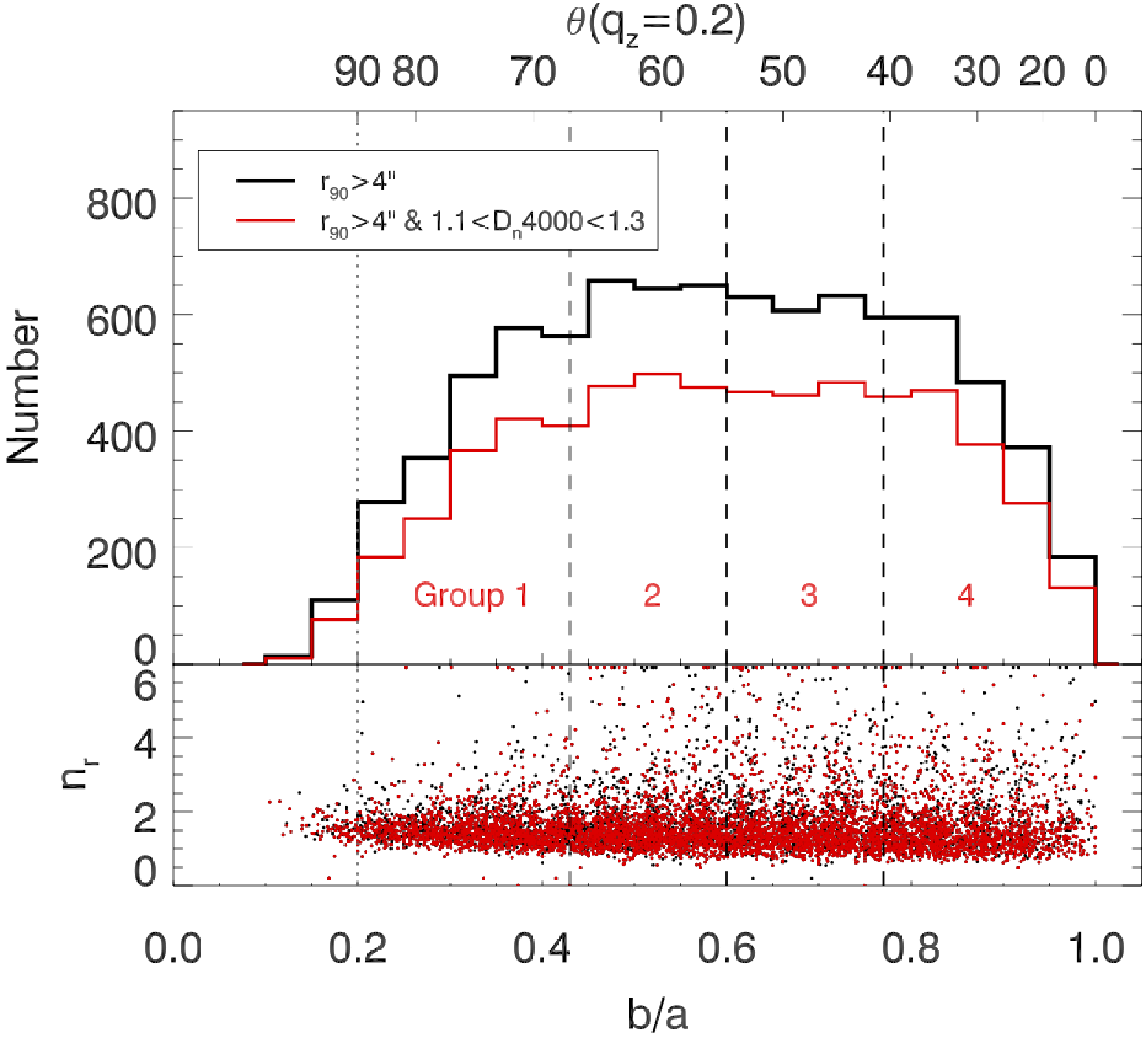} &
\includegraphics[scale=0.5]{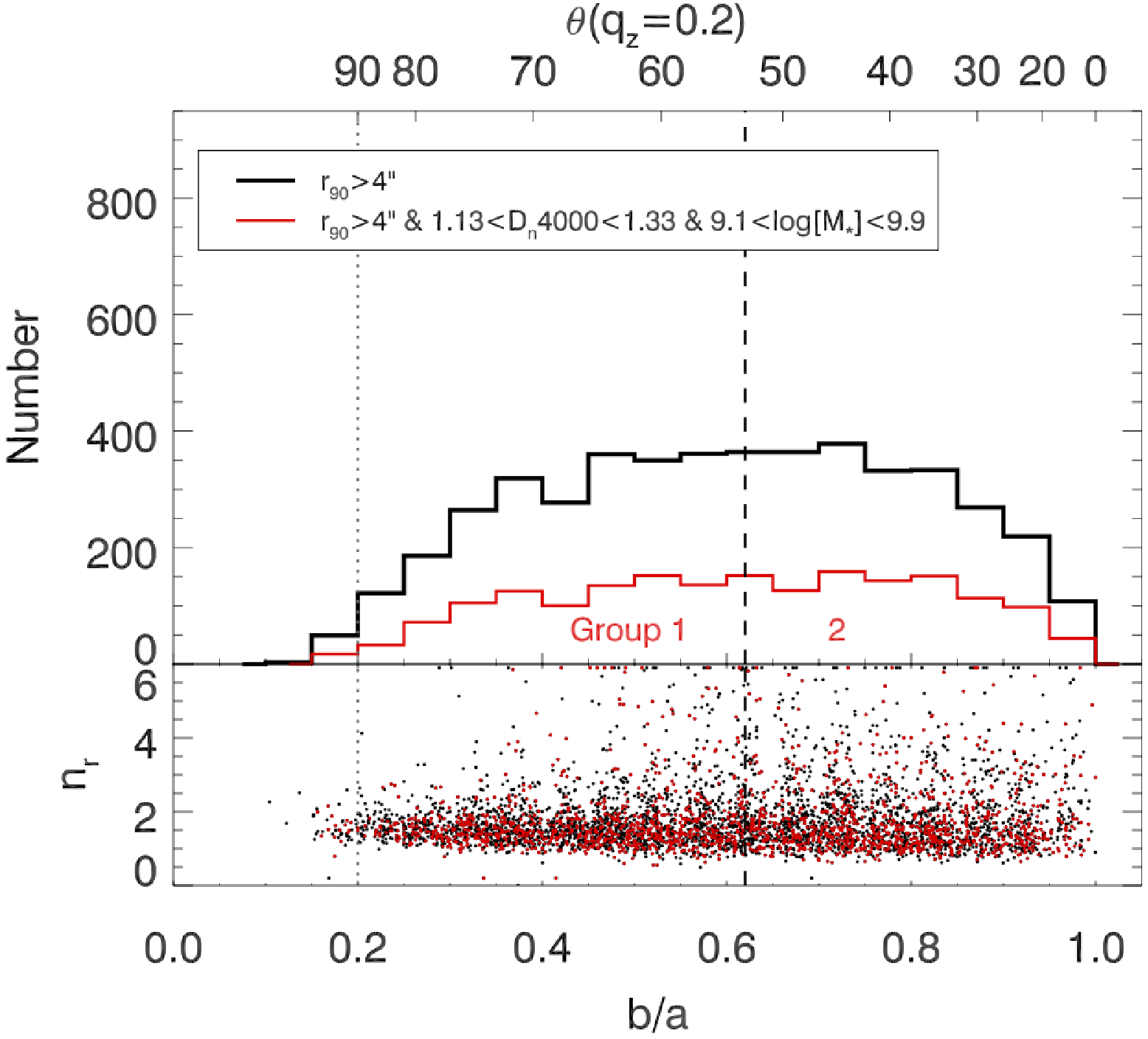} \\
\end{array}$
\end{center}
\vspace{-0.2cm}
\caption{The distribution of axial ratios $b/a$ for galaxies with $r_{90}>4\arcsec$ in our parent (\textit{left}) UV-optical ($N=8441$) and (\textit{right}) UV-NIR ($N=4658$) samples (black lines). For a sample of perfectly thin disk galaxies with random orientation, the expected distribution is flat. However, the intrinsic scale height of disks $q_z>0$ will prevent axial ratios below the scale height from being observed. Similarly, any intrinsic ellipticity in the disk $\epsilon>0$ will tend to reduce the number of cases with very high axial ratios. The top axis shows the galaxy inclination angle assuming $q_z=0.2$ and the dotted line corresponds to a completely edge-on galaxy in this scenario. The red lines show subpopulations that have additional constraints utilized for the analysis in Section~\ref{derive_curve}. The dashed lines separate the groups used to examine the effect of inclination and are chosen to be wide enough such that scale height and ellipticity do not have a substantial impact on the results. The lower panel shows the $r$-band S\'ersic index as a function of the axial ratio, where the full $r_{90}>4\arcsec$ sample is in black and the subpopulation is in red. The majority of the galaxies are disk-like with $n<2.5$.
\label{fig:b2a_hist}}
\end{figure*}

The effects of seeing are also important to consider when attempting to measure inclination. Seeing will cause galaxies with small angular extents to have rounder isophotes than their intrinsic value. We use the radius enclosing 90\% of the Petrosian flux in the $r$ band, $r_{90}$, from the SDSS database to determine the size of each galaxy. To mitigate the issue of seeing, we select only galaxies with $r_{90}>4\arcsec$ for analysis, which corresponds to the minimum size required for reliable spiral classification in Galaxy Zoo \citep{masters10}. We have examined the appearance of numerous galaxies in our sample with $r_{90}<4\arcsec$ and we confirm that $r_{90}=4\arcsec$ is an adequate threshold for reliable inclination determinations. Imposing a cut of $r_{90}>4\arcsec$ leaves us with a sample of 8441 galaxies (i.e., 14\% of the parent sample lies below this threshold) for our UV-optical sample and 4658 for our UV-NIR sample. The distribution of axial ratios $b/a$ for this sample of galaxies is shown in Figure~\ref{fig:b2a_hist}. For reference, we also show the corresponding inclination values assuming a value of $q_z=0.2$ for the vertical scale height.

For subsequent analysis we separate the UV-optical sample into the four groups (subpopulations) and the UV-NIR sample into two groups that are demonstrated in Figure~\ref{fig:b2a_hist}. The groups are chosen to be wide enough such that the scale height, which would impact our ability to select the most edge-on cases, and ellipticity, which would impact our ability to select the most face-on cases, do not have a substantial effect on our results. This is because all nearly all edge-on and face-on cases should still reside in the lowest and highest axial ratio group, respectively, for typical values of $q_z$ and $\epsilon$.

As a check on influence of the S\'ersic index, which may influence the reliability of axial ratios as an indicator of inclination, we explored the effects of additionally restricting our sample to only galaxies of lower S\'ersic index (i.e., a more disk-like profile). In particular, when we select samples with only $n<2.5$ or $n<1.5$ for the analysis presented in Section~\ref{derive_curve}, we find that the results do not appear to show any significant differences relative to those found without imposing this restriction, albeit at higher uncertainty due to the smaller sample size. Therefore, we feel confident that our results are not strongly dependent on the S\'ersic index.

\subsection{Characterizing Attenuation}\label{method_attenuation}
The dust attenuation in each galaxy is quantified through the Balmer decrement, $F(\mathrm{H}\alpha)/F(\mathrm{H}\beta)$, where H$\alpha$ and H$\beta$ are located at 6562.8~\AA\ and 4861.4~\AA, respectively. Following \citet{calzetti94}, we define the Balmer optical depth as
\begin{equation}\label{eq:tau}
\tau_B^l = \tau_{\mathrm{H}\beta} - \tau_{\mathrm{H}\alpha} = \ln \left(\frac{F(\mathrm{H}\alpha)/F(\mathrm{H}\beta)}{2.86}\right)\,,
\end{equation}
where the value of 2.86 comes from the theoretical value expected for the unreddened ratio of $F(\mathrm{H}\alpha)/F(\mathrm{H}\beta)$ undergoing Case B recombination with $T_{\mathrm{e}}=10^4$~K and $n_{\mathrm{e}}=100$~cm$^{-3}$ \citep{osterbrock89,osterbrock&ferland06}. The superscript \textit{l} is used to emphasize that this quantity is coming from emission lines and should be distinguished from optical depths associated with the stellar continuum. 

If one assumes knowledge of the total-to-selective extinction, $k(\lambda)\equiv A_\lambda /E(B-V)$, then $\tau_B^l$ can be directly related to the color excess of the nebular gas, $E(B-V)_{\mathrm{gas}}$, through
\begin{equation}\label{eq:EBV_gas}
E(B-V)_{\mathrm{gas}}=\frac{1.086\tau_B^l}{k(\mathrm{H}\beta)-k(\mathrm{H}\alpha)} \,,
\end{equation}
where $A(\lambda)$ is the total extinction at a given wavelength. For reference, the MW has a value of $k(\mathrm{H}\alpha)-k(\mathrm{H}\beta)=1.257$ \citep{fitzpatrick99}.

\begin{figure}
\begin{center}
\includegraphics[scale=0.5]{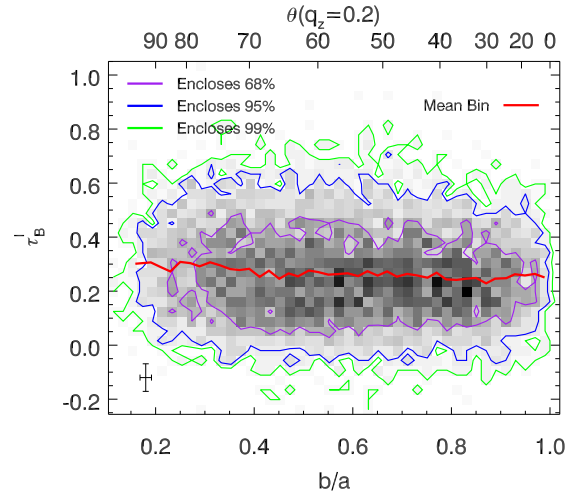}
\end{center}
\vspace{-0.2cm}
\caption{The Balmer optical depth $\tau_B^l$ as a function of axial ratio $b/a$ for the UV-optical sample. The range and average value of $\tau_B^l$ does not vary strongly with $b/a$, although a slight increase in $\tau_B^l$ at lower $b/a$ is evident as seen in previous studies \citep[e.g.,][]{yip10,chevallard13}. For reference, if we assume the ionized gas experiences an extinction curve similar to the MW ($R_V=3.1$), the values of $\tau_B^l=0.4$ and 0.6 ($\sim$68\% and 95\% upper boundary) correspond to $A_{V,\mathrm{gas}}=1.07$ and 1.61, respectively (using equation~\ref{eq:EBV_gas}).
\label{fig:tau_vs_b2a}}
\end{figure}

We show the values of $\tau_B^l$ as a function of $b/a$ for our sample in Figure~\ref{fig:tau_vs_b2a}. Interestingly, it can be seen that the range and average value of $\tau_B^l$ does not vary strongly with $b/a$, although there may be a slight increase in $\tau_B^l$ at lower $b/a$. Similar results have been seen in previous studies \citep[e.g.,][]{yip10,chevallard13,catalan-torrecilla15} and also predicted theoretically \citep[e.g.,][]{tuffs04,chevallard13}. Two plausible explanations for the mild trend of $\tau_B^l$ with inclination are presented in \citet{wild11}. First, these emission lines may undergo the majority of their total attenuation from dust in their surrounding star-forming region, such that any additional attenuation from the diffuse dust component at high inclinations would have only a minor influence. Second, at high inclinations the line of sight to star-forming regions may become optically thick such that only those lying on the outer skin can be viewed and these cases would be subject to only modest additional attenuation from the diffuse dust component. There is evidence against the second scenario being the primary mechanism for the lack of a trend due to the excellent agreement between the total attenuation (or SFRs) inferred from Balmer decrements (Balmer-corrected H$\alpha$) and those based on flux ratios of IR emission and observed H$\alpha$ (IR+observed H$\alpha$) \citep{calzetti07, kennicutt09} regardless of galaxy inclination \citep{prescott07, catalan-torrecilla15}. The first scenario is favored theoretically by \citet{chevallard13}, in which their radiative transfer modeling produces total effective optical depths (i.e., that measured in attenuation) from the ISM+star-forming regions that are typically less than unity at optical wavelengths. We will discuss this further in Section~\ref{lit_compare}.

For galaxies where the UV flux density is dominated by recent star formation (i.e., small contribution from older stars), the intrinsic UV spectral slope has an expected value of $\beta_0\sim-2.1$ \citep{calzetti00}. Therefore, the effects of dust attenuation are also often inferred from the observed UV spectral slope $\beta$, where
\begin{equation}
F(\lambda)\propto\lambda^\beta\,,
\end{equation}
and $F(\lambda)$ is the flux density in the range $1250\le\lambda\le2600$~\AA.

For this study, we make use of the UV power-law index $\beta_{\rm{GLX}}$ measured from observed \textit{GALEX} FUV and NUV photometry,
\begin{equation}
\beta_{\rm{GLX}} = \frac{\log[F_\lambda(\mathrm{FUV})/F_\lambda(\mathrm{NUV})]}{\log[\lambda_{\mathrm{FUV}}/\lambda_{\mathrm{NUV}}]}\,,
\end{equation}
where the flux density is in erg~s$^{-1}$~cm$^{-2}$~\AA$^{-1}$, $\lambda_{\rm{FUV}}=1516\ang$, and $\lambda_{\rm{NUV}}=2267\ang$. There is no need to perform $k$-corrections on the flux density if one assumes a power-law fit to the region described above. However, this value can be influenced by the presence of a 2175~\AA\ absorption feature (for low redshifts), which will tend to steepen the slope (higher negative values) due to a decrease in the NUV flux density \citep[see][]{battisti16}.

\section{Deriving the Dust Attenuation Curves}\label{derive_curve}
\subsection{Galaxy Templates}\label{templates}
To empirically determine our attenuation curves, we follow the same methodology as previously employed in \citet{battisti16,battisti17}. This requires comparing the SEDs of galaxies with differing amounts of dust attenuation while also accounting for intrinsic differences between the stellar populations among our sample, which can lead to significant SED variation independent of the total level of dust attenuation.  Following previous studies of this kind, we utilize ensembles of many galaxies to construct templates with similar ``average'' stellar populations \citep{calzetti94,wild11,reddy15}. Another issue is the well known age/dust degeneracy in which the reddening of a young, dusty stellar population can be similar to that of an old, dust-free population \citep[e.g.,][]{witt92,gordon97}. Due to this effect, the inferred attenuation curve can be subject to systematic errors if intrinsic differences exist among the average SEDs as a function of the relative attenuation. For example, SFGs experiencing more dust attenuation also, on average, tend have slightly older average stellar population ages (likely linked to the production of metals/dust from older stars) that causes their SEDs to be intrinsically redder than less attenuated galaxies \citep{battisti16}. This effect, if unaccounted for, would lead to a steeper inferred attenuation curve. In order to mitigate these issues, we select a sample of galaxies with roughly similar SFHs and stellar populations based on the strength of the 4000~\AA\ break feature \citep[$D_n4000$;][]{kauffmann03a}.

For our parent sample, a linear relationship is observed between $\beta_{\rm{GLX}}$ and $\tau_B^l$, although with a very large dispersion \citep{battisti16}. Similar relationships are observed for local starburst galaxies \citep{calzetti94} and also in SFGs at $z\sim2$ \citep{reddy15}. For the purpose of this analysis, we assume a foreground-like dust component is dominating the attenuation, which is consistent with the linear relationship between these quantities, and provides us with a straightforward manner to derive the effective attenuation curve. However, we stress that more complex geometries are likely to occur and may account for some of the scatter in the observed relationship \citep[see][for more discussion]{battisti16,battisti17}.

In the scenario in which a foreground-like dust component is dominating the attenuation, the optical depth is expected to behave according to
\begin{equation}
\tau_{n,r}(\lambda) = -\ln \frac{F_n(\lambda)}{F_r(\lambda)} \,,
\end{equation}
where $\tau_{n,r}$ corresponds to the dust optical depth of template $n$, with flux density $F_n(\lambda)$, relative to a reference template $r$, with flux density $F_r(\lambda)$, and it is required that $n>r$ for comparison. From this relation, it is possible to determine the selective attenuation, $Q_{n,r}(\lambda)$, 
\begin{equation}\label{eq:Q_def}
Q_{n,r}(\lambda) = \frac{\tau_{n,r}(\lambda)}{\delta \tau_{Bn,r}^l} \,,
\end{equation}
where $\delta \tau_{Bn,r}^l=\tau_{Bn}^l-\tau_{Br}^l$ is the difference between the Balmer optical depth of template $n$ and $r$. The normalization for the selective attenuation is arbitrary. Following our previous work, we select $Q_{n,r}(5500\mathrm{\AA})=0$ as the zero-point.

\begin{figure*}
\begin{center}$
\begin{array}{cc}
\includegraphics[scale=0.5]{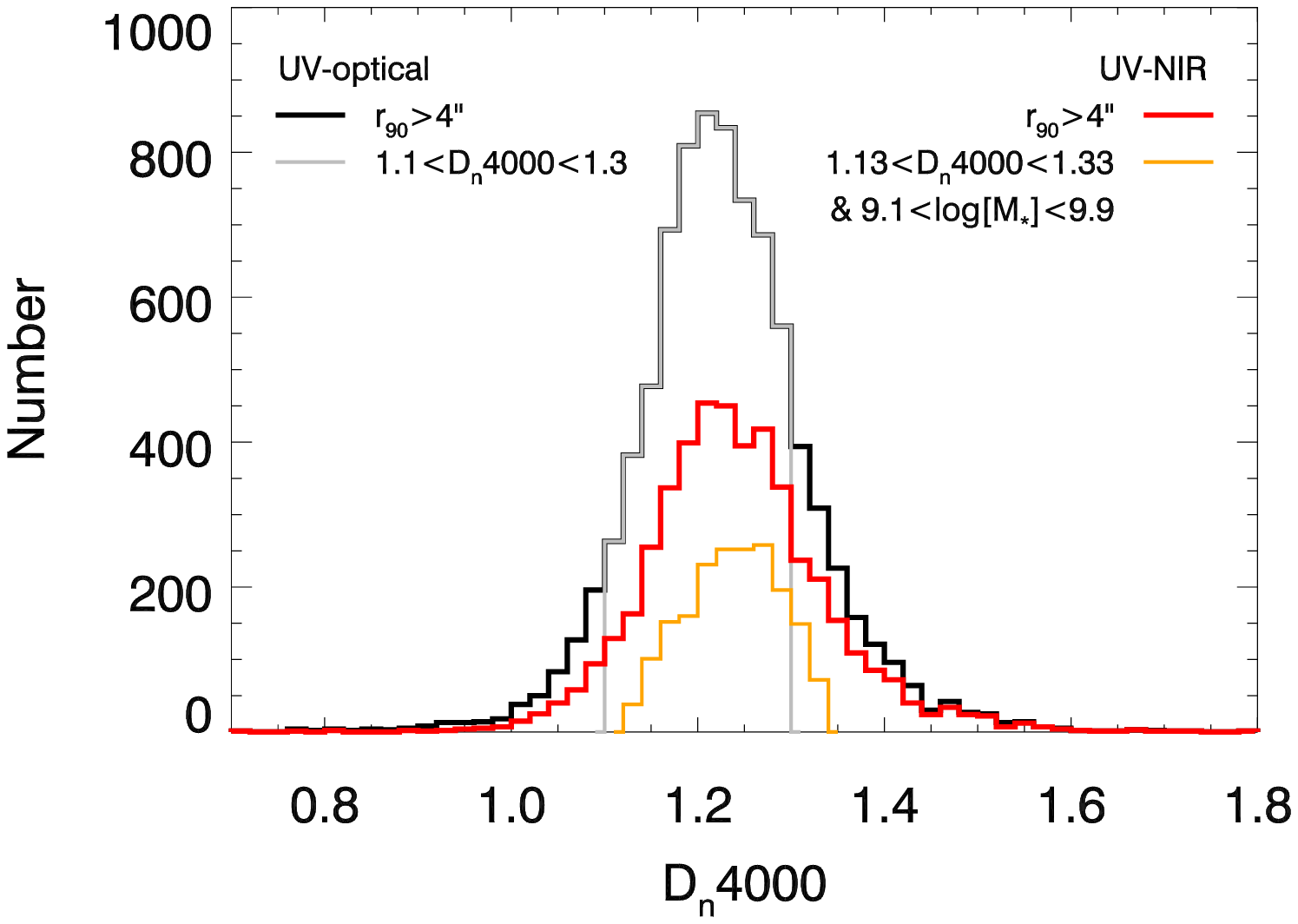} &
\includegraphics[scale=0.5]{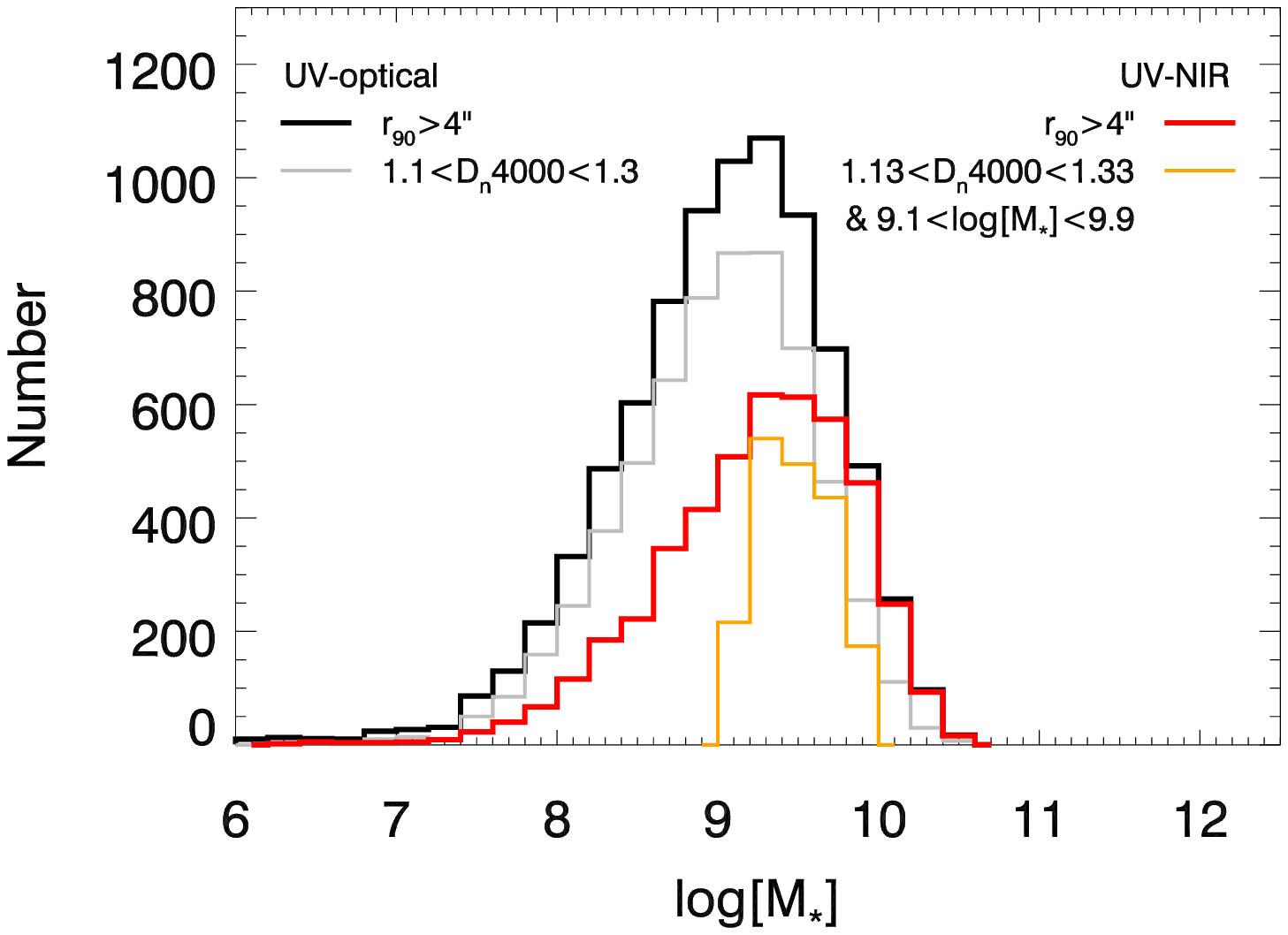} \\
\end{array}$
\end{center}
\vspace{-0.2cm}
\caption{The distribution of (\textit{left}) $D_n4000$ and (\textit{right}) $M_*$ values for galaxies with $r_{90}>4\arcsec$ in our parent UV-optical (black lines) and UV-NIR (red lines) samples. The gray and orange lines show subpopulations that have additional constraints utilized for the analysis in Section~\ref{derive_curve} and are chosen to retain the largest fraction of the galaxy sample while also reducing the influence of stellar population age on the derived attenuation curve. \label{fig:Dn4000_gm_hist}}
\end{figure*}

In order to construct the templates, we select galaxies in a narrow range of $D_n4000$ values for comparison. This selection removes the systematic trend seen between dustier galaxies having slightly larger $D_n4000$ values (older average stellar population age), which impacts the inferred attenuation in the UV-optical wavelength range \citep{battisti16}. We select galaxies within a window of $\Delta D_n4000=0.2$, which is centered on the mean value of 1.2 and 1.23 for the UV-optical and UV-NIR samples, respectively. For the NIR analysis, we also impose a restriction on the stellar mass range of $9.1<\log[M_* (M_{\odot})]<9.9$ to mitigate the variation in the NIR due to the old, low-mass stellar populations \citep{battisti17}, for which $D_n4000$ is insufficient to constrain because it is primarily linked to intermediate mass stars. The logic behind the chosen $D_n4000$ and $M_*$ windows is to retain the largest fraction of the galaxy sample while also reducing the influence of stellar population age on the derived attenuation curve. We find that using windows narrower than those mentioned above does not lead to significant differences in the resulting attenuation curve \citep[relative to the dispersion, which increases with narrower windows; see][]{battisti16, battisti17}. For reference, we show the histograms of the $D_n4000$ and $M_*$ values for these subpopulations relative to the $r_{90}>4\arcsec$ parent samples in Figure~\ref{fig:Dn4000_gm_hist}. These criteria provide us with samples of 6289 and 1860 galaxies for the UV-optical and UV-NIR analysis, respectively. 

\begin{figure*}
\begin{center} 
$\begin{array}{cc}
\includegraphics[scale=0.5]{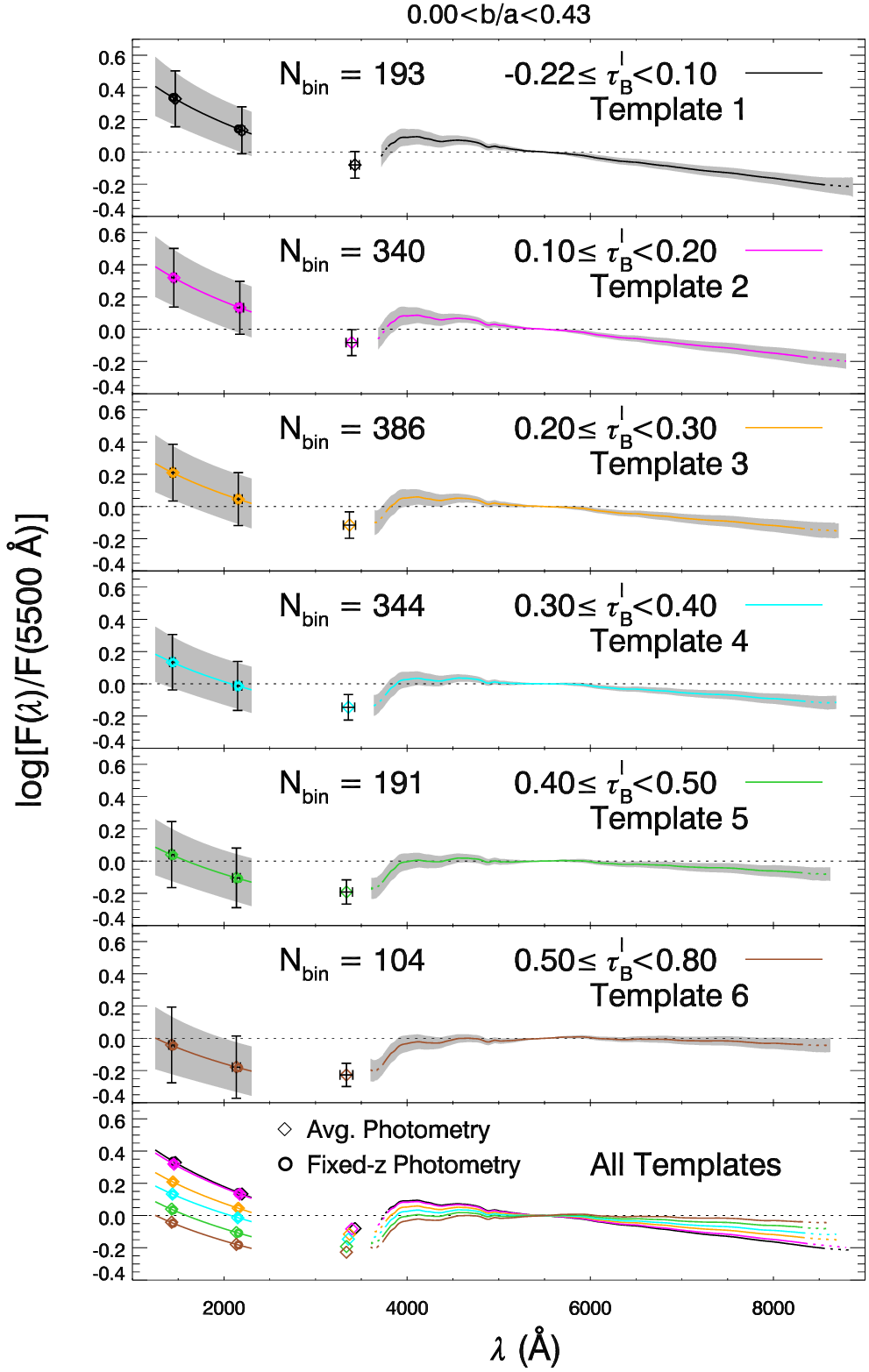} & \hspace{1cm}
\includegraphics[scale=0.5]{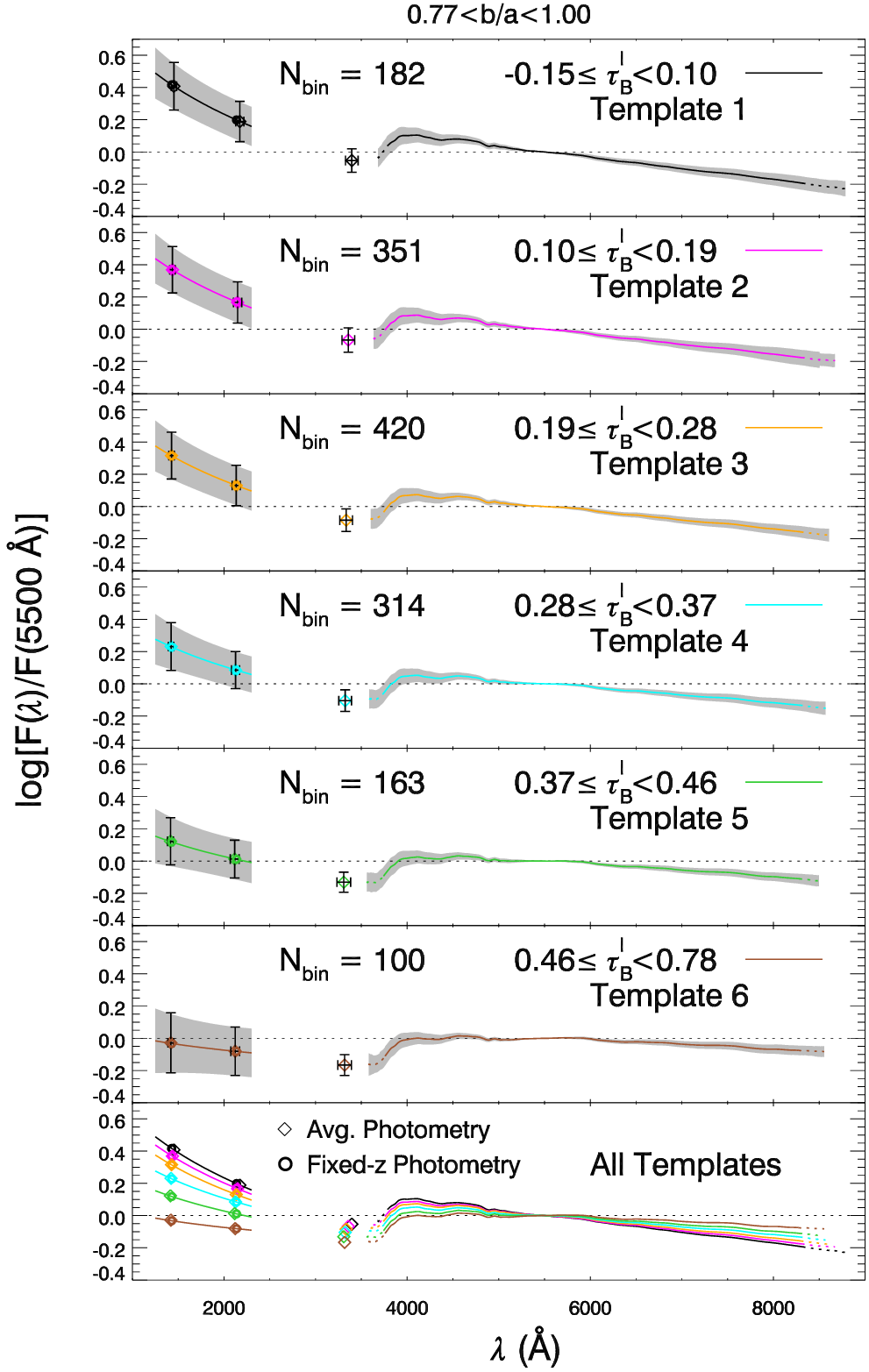} \\
\end{array}$
\end{center}
\vspace{-0.2cm}
\caption{Average UV-optical flux density, normalized at 5500~\AA, within bins of $\tau_B^l$ for galaxies separated by axial ratio (proxy for inclination) and with $1.1<D_n4000<1.3$. The \textit{left} shows the low axial ratio sample (higher inclination angle, closer to edge-on) with $0.00<b/a<0.43$ ($N=1558$) and the \textit{right} shows the high axial ratio sample (lower inclination angle, closer to face-on) with $0.77<b/a<1.00$ ($N=1530$). The range in $\tau_B^l$ and the number of sources in each bin, $N_{\mathrm{bin}}$, are shown in each panel. The flux density shown for the region of $1250<\lambda<2300$~\AA\ is based on the average \textit{GALEX} FUV and NUV data assuming it follows a power-law $F(\lambda)\propto\lambda^\beta_{\rm{GLX}}$. The optical measurements are from SDSS spectroscopy. The gray regions denote the area enclosing approximately 68\% of the population. The dotted regions in the optical spectra indicate the average obtained from less than the full sample in that bin (due to varying redshifts), but still containing $>50\%$ of the bin sample. The diamonds show the average flux densities for \textit{GALEX} FUV, NUV, and SDSS $u$ photometry, with errorbars denoting the 1$\sigma$ range in rest-frame wavelength and flux density values spanned in each bin, respectively. The bottom panel shows a comparison of the average flux density of each bin without the dispersion included. It can be seen that the shape of the UV region differs with axial ratio. \label{fig:F_vs_lam}}
\end{figure*}

To examine the effects of inclination we separate our sample according to axial ratio and adopt 4 groups for the UV-optical sample, as shown in Figure~\ref{fig:b2a_hist}. The galaxies within each subpopulation are separated into 6 bins of $\tau_B^l$, spaced by $\Delta\tau_B^l\sim0.1$ (maintaining $\gtrsim$100 sources in each bin), and the average SED is determined. We refer the reader to \citet{battisti16} for a full description of the techniques used for averaging the UV and optical data. For the analysis here we will also retain the average FUV and NUV photometry, in addition to the power-law fit for the UV from $1250<\lambda<2300$~\AA\, because we will also consider scenarios with a 2175~\AA\ feature where the shape of the UV attenuation would deviate from this form. To account for the slight variation in the average redshift for each bin, we also adopt a fixed effective wavelength corresponding to the average redshift of each subpopulation, typically $\bar{z}\sim0.06$, such that the selective attenuation for each band is determined in a consistent manner. We use the averaged UV power-law to determine these fixed-$z$ flux density values, but note that their difference relative to the averaged photometry is small such that the assumption has minimal importance. The average flux density templates for the lowest and highest axial ratio UV-optical subpopulations can be seen in Figure~\ref{fig:F_vs_lam}. Comparing the lowest panel of these two cases (showing the templates with different $\tau_B^l$ relative to one another), it can be seen that the change in UV shape of the average templates with increasing $\tau_B^l$ are noticeably different, with the low axial ratio (higher inclination) galaxies retaining a bluer slope at higher Balmer optical depths.

\begin{figure*}
\begin{center} $
\begin{array}{cc}
\includegraphics[scale=0.5]{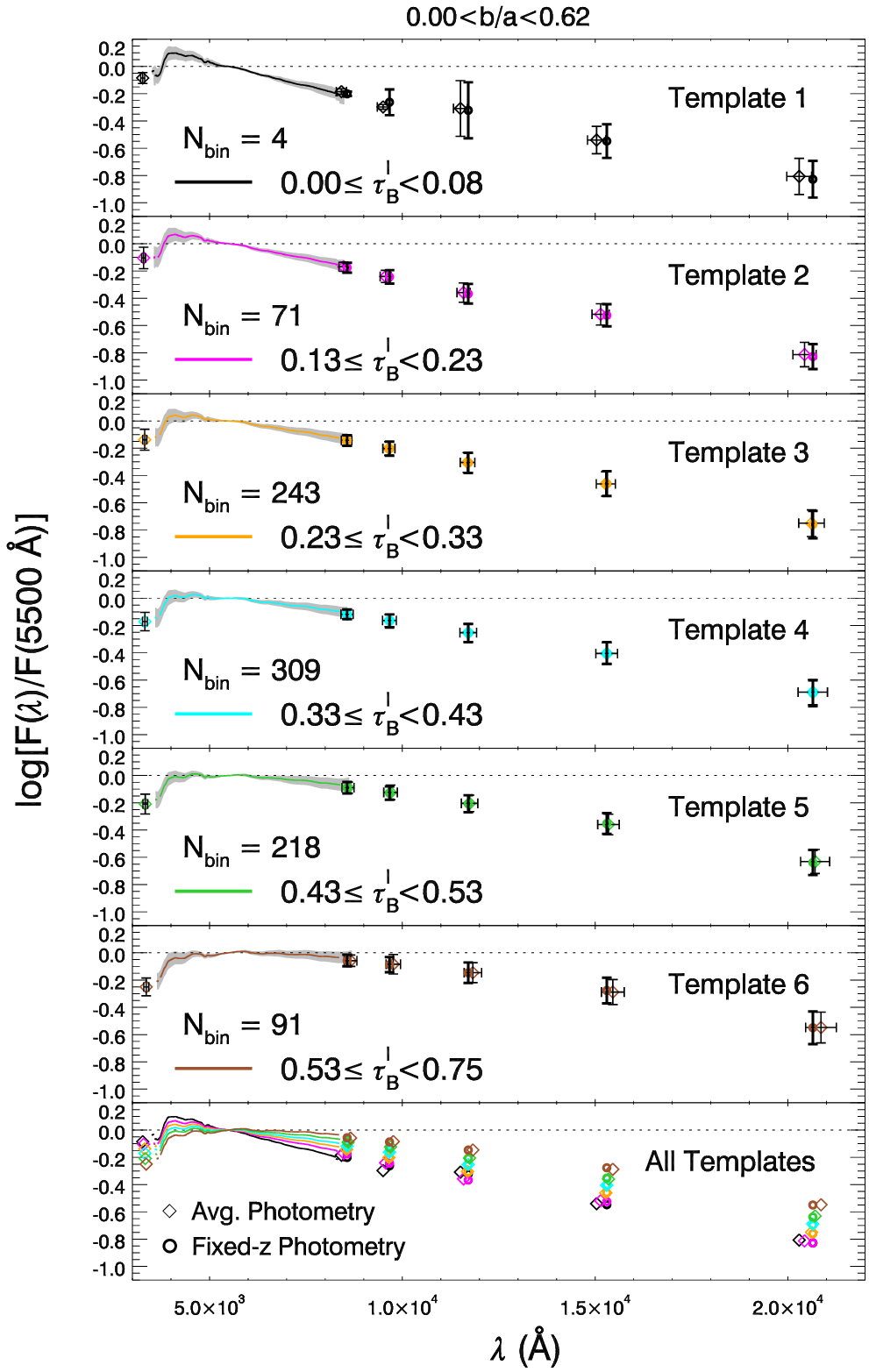} & \hspace{1cm}
\includegraphics[scale=0.5]{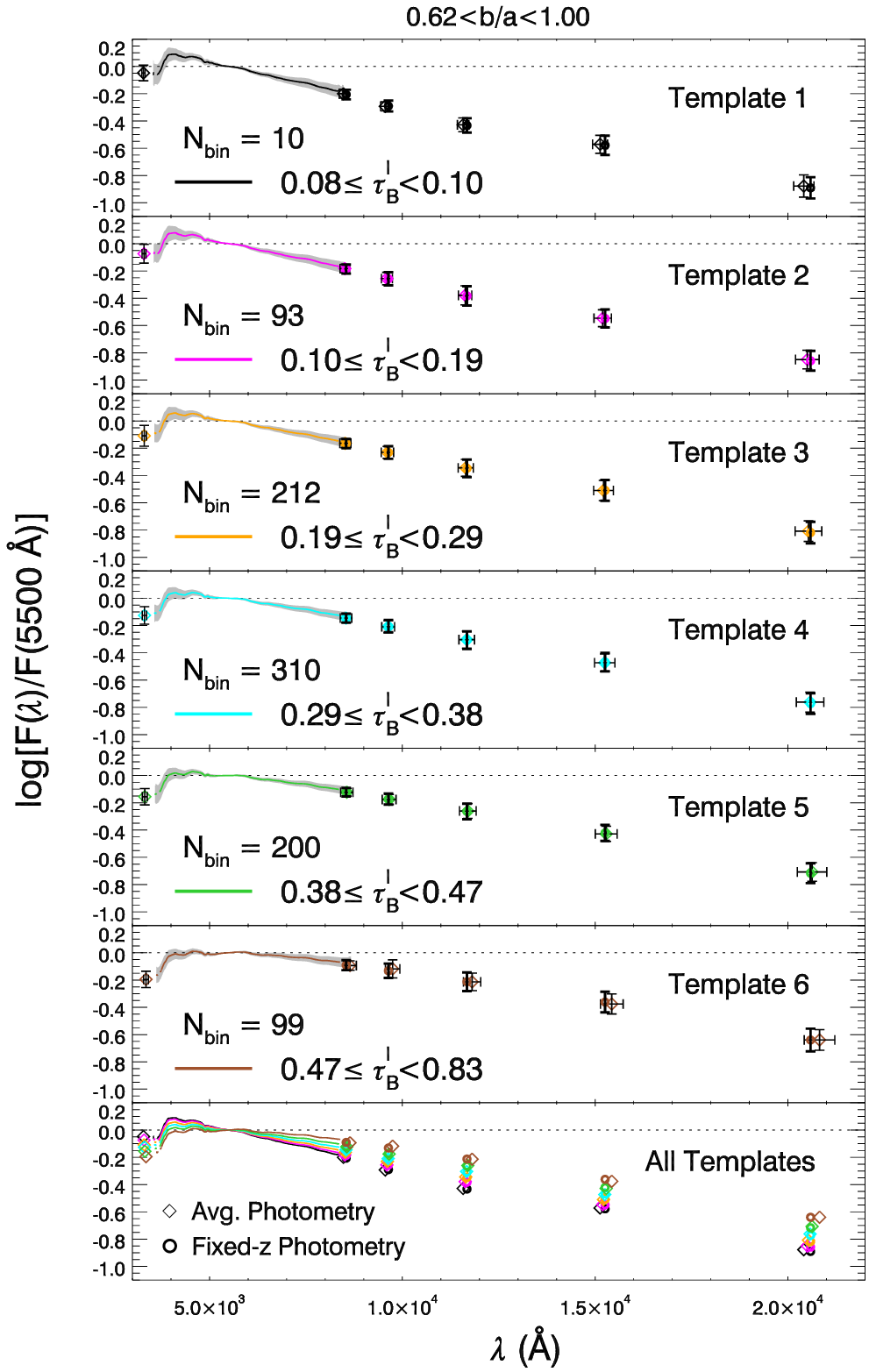} \\
\end{array}$
\end{center}
\vspace{-0.2cm}
\caption{Average optical-NIR flux density for galaxies separated by axial ratio and with $1.13<D_n4000<1.33$ \& $9.1<\log[M_* (M_{\odot})]<9.9$.  The \textit{left} shows the low axial ratio sample with $0.00<b/a<0.62$ ($N=936$) and the \textit{right} shows the high axial ratio sample with $0.62<b/a<1.00$ ($N=924$). Lines and symbols are the same as Figure~\ref{fig:F_vs_lam}. There are no obvious differences in the shape of the templates with inclination.
 \label{fig:F_vs_lam_IR}}
\end{figure*}

For the smaller UV-NIR sample we adopt only 2 groups, as illustrated in Figure~\ref{fig:b2a_hist}. We follow the methodology outlined in \citet{battisti17}, for which we interpolate the average flux density of the observed SDSS $z$ and UKIDSS $Y$, $J$, $H$, and $K$ bands (in log($F_\lambda$)-log($\lambda$) space) to a fixed wavelength (the effective wavelength for $\bar{z}$) for determining the selective attenuation at each band. The behavior of the NIR spectral region is relatively smooth such that interpolating values for small wavelength offsets is reasonable. The average flux density templates for the UV-NIR subpopulations can be seen in Figure~\ref{fig:F_vs_lam_IR}. Comparing the lowest panels of this Figure, it can be seen that there are no obvious differences in the shape of the templates with increasing $\tau_B^l$ at optical and NIR wavelengths for the different inclination groups.

\begin{figure*}
\begin{center} $
\begin{array}{cc}
\includegraphics[scale=0.5]{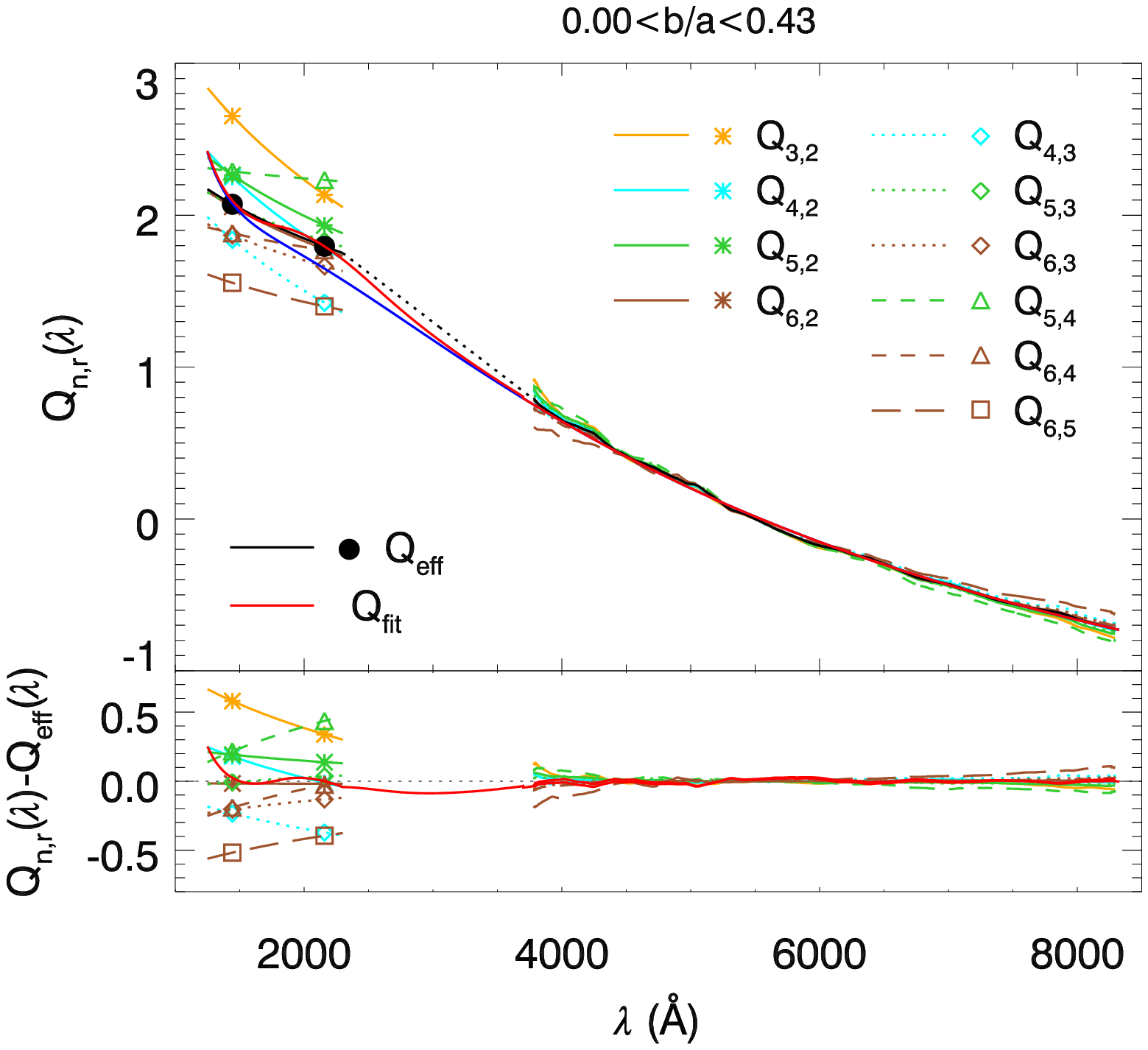} &
\includegraphics[scale=0.5]{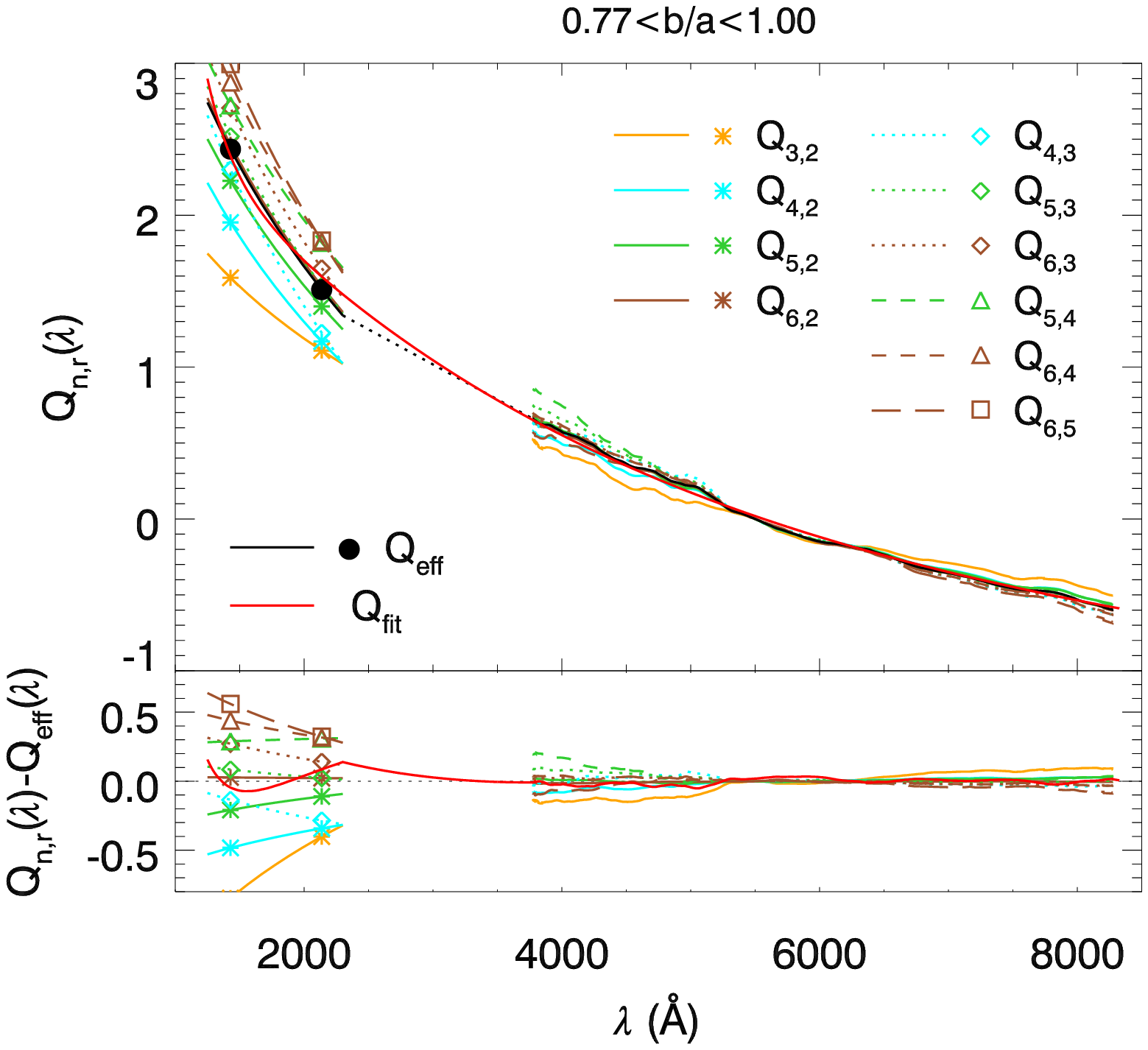} \\
\end{array}$
\end{center}
\vspace{-0.2cm}
\caption{UV-optical selective attenuation curve, $Q_{n,r}(\lambda)$, for our axial ratio subpopulations based on comparing a given template, $n$, to a reference template, $r$, at lower $\tau_B^l$. Also shown is the effective curve, $Q_{\mathrm{eff}}(\lambda)$ (solid black line and circles), which is the average value of $Q_{n,r}(\lambda)$ for the cases shown. The gap region between 2300~\AA\ and 3750~\AA\ is denoted with a dotted line corresponding to a linear interpolation between the end points and is not used for constraining the fit. The solid red line is our best fit to $Q_{\mathrm{eff}}(\lambda)$ (see Section~\ref{bump_fit}). For the low axial ratio sample (\textit{left}), the NUV point has an excess relative to a baseline constructed from a fit to only the FUV and optical data (solid blue line), which we interpret to be due to a 2175~\AA\ feature. The lower panel shows the residual between each curve relative to $Q_{\mathrm{eff}}(\lambda)$. \label{fig:Q_eff}}
\end{figure*}

\begin{figure*}
\begin{center}$
\begin{array}{cc}
\includegraphics[scale=0.5]{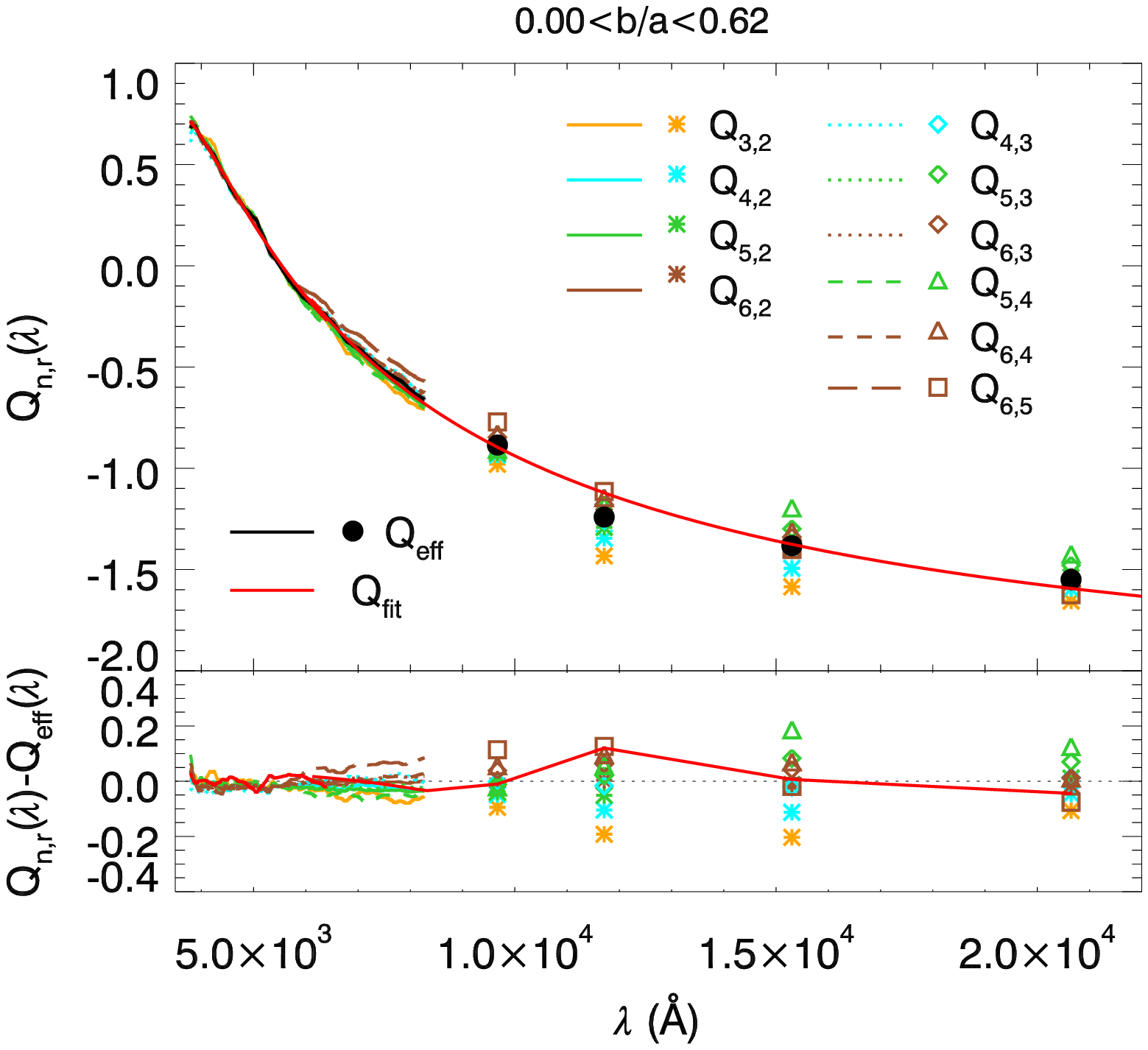} &
\includegraphics[scale=0.5]{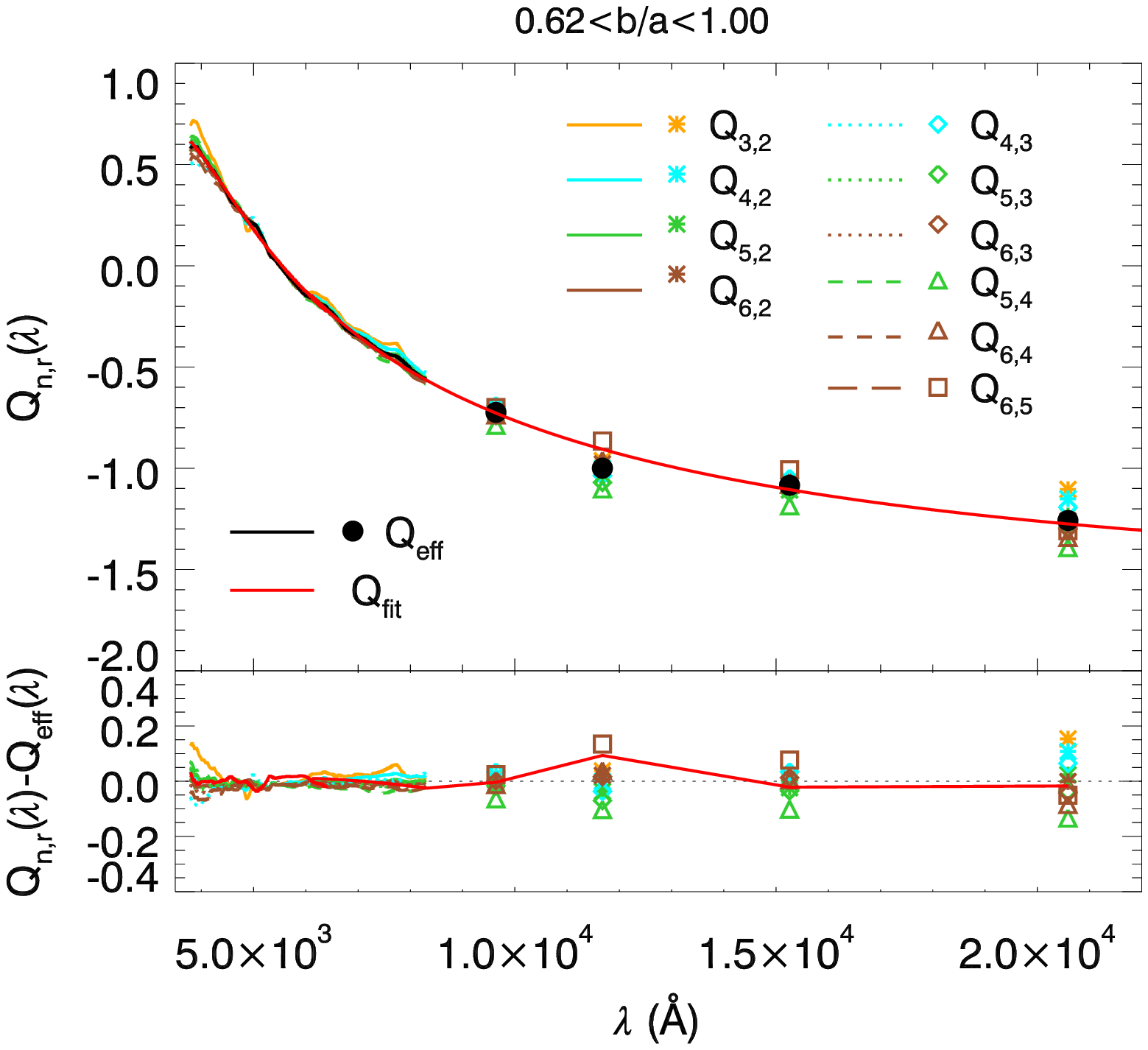} \\
\end{array}$
\end{center}
\vspace{-0.2cm}
\caption{Optical-NIR selective attenuation curve, $Q_{n,r}(\lambda)$, for our axial ratio subpopulations. Lines and symbols are the same as Figure~\ref{fig:Q_eff}. The fitting methodology is described in Section~\ref{bump_fit}. \label{fig:Q_eff_IR}}
\end{figure*}

We show the selective attenuation curves that are obtained from the lowest and highest axial ratio UV-optical templates in Figure~\ref{fig:Q_eff}. We exclude the use of template 1 in our analysis because this average SED is nearly identical to template 2, which we attribute to the intrinsic variation being dominant over the effects of attenuation at these low values of $\tau_B^l$. It can be seen in Figure~\ref{fig:Q_eff} that templates 2-6 give similar selective attenuation curves, implying that adopting a single selective attenuation curve is reasonable to characterize each subpopulation. We determine the effective attenuation curve, $Q_{\mathrm{eff}}(\lambda)$, by taking the average value of $Q_{n,r}(\lambda)$ found from templates 2-6. We adopt the same approach for the UV-NIR subpopulations, which are shown in Figure~\ref{fig:Q_eff_IR}. In the next section we outline the method used to fit the shape of these selective attenuation curves.

\subsection{Fitting the Selective Attenuation and Bump Feature}\label{bump_fit}
In this section we outline the method for fitting the average selective attenuation curves of our subpopulations. These methods are identical to those presented in \cite{battisti16}, except that we also consider the possibility of an additional MW-like 2175\AA\ feature of variable bump strength. This feature is assumed to account for the NUV excess in $Q(\lambda)$ for the low $b/a$ subpopulation (higher inclination, Figure~\ref{fig:Q_eff}, left). First, we fit the optical region of $Q(\lambda)$ from the SDSS spectroscopic data together with the average FUV value using a third order polynomial, which acts as our baseline curve. The excess of the NUV $Q(\lambda)$ value relative to this baseline is then assumed to be a MW-like 2175\AA\ feature. It is worth noting that the inferred bump strength is directly dependent on the assumed baseline, which introduces additional uncertainty into the method. The MW bump feature is well characterized using a Lorentzian-like Drude profile \citep{fitzpatrick&massa90}
\begin{equation}
D(\lambda) = \frac{E_b(\lambda\,\Delta\lambda)^2}{(\lambda^2-\lambda_0^2)^2+(\lambda\,\Delta\lambda)^2} \,,
\end{equation}
where $\lambda_0=2175$~\AA\ is the wavelength of the feature, $\Delta\lambda$ is its FWHM, and $E_b$ is the bump intensity. The average MW extinction curve has values of $\Delta\lambda\sim470$~\AA\ and $E_b\sim3.30$ \citep{fitzpatrick99}.

\begin{figure}
\begin{center}
\includegraphics[scale=0.5]{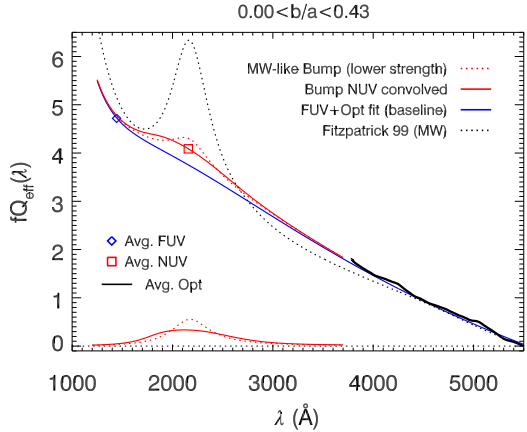} 
\end{center}
\vspace{-0.2cm}
\caption{Demonstration of our 2175\AA\ feature fitting method for the high-inclination subpopulation. The baseline is determined by fitting the FUV+optical data to a third-order polynomial. Assuming that the NUV excess relative to this baseline is due to a bump with MW-like properties ($\lambda_0=2175$~\AA\ and $\Delta\lambda=470$~\AA) requires a bump strength of $E_b=0.55$, or about 17\% the strength of the MW (see Section~\ref{bump_fit} for different assumptions). The MW curve is also shown for reference \citep{fitzpatrick99}. \label{fig:Q_eff_feature_fit}}
\end{figure}

To fit the residual NUV value, we use a Drude profile with  $\lambda_0=2175$~\AA\ and $\Delta\lambda=470$~\AA\ that is convolved with the \textit{GALEX} NUV filter. This convolution acts to suppress the apparent strength of the feature \citep[e.g.,][]{battisti16}. The corresponding strength of the bump required to fit the NUV is $E_b=0.55$, or about 17\% the strength of the MW. We illustrate the fit in Figure~\ref{fig:Q_eff_feature_fit}, where this fit includes the differential reddening factor $f$ (see Section~\ref{attenuation curve}) on $Q(\lambda)$ so that the MW extinction curve can be shown for comparison. However, this value is dependent on the assumed value for $\lambda_0$ and $\Delta\lambda$. Adopting values from \citet{noll09a} for $z\sim2$ galaxies, which tend to have higher $\lambda_0$ and lower $\Delta\lambda$ than the MW values (see their table2), requires bump strengths ranging from $0.70<E_b<0.87$ or 21-26\% the MW value. Given that we do not have adequate information to constrain $\lambda_0$ and $\Delta\lambda$, we choose to adopt the fit using the MW values for the remainder of the paper. Regardless of the adopted parameters, the assumed bump is substantially lower than that of the average MW curve or the LMC2 supershell and are consistent with the bump strengths found in \citet{noll09a} ($0.47<E_b<0.93$).  

Performing a similar analysis on the other $b/a$ bins results in a ``negative feature'' being preferred and therefore it is assumed that a feature is not required. For these cases, we adopt our standard approach and fit the UV-optical values of $Q_{\mathrm{eff}}(\lambda)$, where the UV is from the average power-law fit, to a single third-order polynomial as a function of $x=1/\lambda$~($\micron^{-1}$):
\begin{multline}\label{eq:Q_fit}
Q_{\mathrm{fit}}(x) = p_0+p_1x+p_2x^2+p_3x^3\,,\\ 0.125\micron\le\lambda<0.832\micron \,.
\end{multline}
The parameters of all fits are outlined in Table~\ref{Tab:Q_vs_param}. 

For the UV-NIR subpopulations, we follow the methodology of \cite{battisti17} and fit the NIR region of $Q_{\mathrm{eff}}(\lambda)$ using the SDSS data longward of 6000~\AA\ combined with the NIR bands to a single second-order polynomial as a function of $x=1/\lambda$~($\micron^{-1}$):
\begin{multline}\label{eq:Q_fit_IR}
Q_{\mathrm{fit}}(x) = p_0+p_1x+p_2x^2\,,\\ 0.63~\micron\le\lambda<2.1~\micron \,. 
\end{multline}
A second-order fit was chosen for the NIR because of the limited data available at these wavelengths. We separately fit the SDSS region ($3750-8325$~\AA), using a third-order polynomial, for determining the differential reddening factor ($f$; Section~\ref{attenuation curve}). The parameters of these fits are outlined in Table~\ref{Tab:Q_IR_vs_param}. Looking at Figure~\ref{fig:Q_eff_IR}, it is apparent that these fits may simplify the shape of the attenuation curve (e.g., slight offset of $J$-band value of $Q_{\mathrm{eff}}$), although we also caution that discrepancies between the photometric values could arise from SED variation that not accounted for using our current selection criteria.

We also examined the effect of adopting the same bins for the UV-optical sample as for the UV-NIR sample (i.e., use only two groups) and find that it results in attenuation curves that are effectively an average of the two lower ($b/a<0.60$) and upper ($b/a>0.60$) groups. This demonstrates that the mass-selection has minimal influence on the derived curve shape at shorter wavelengths. Similarly, we find the UV behavior for the attenuation curve inferred from our UV-NIR sample is consistent with the larger UV-optical sample, although with a larger dispersion that we attribute to the smaller sample size.

\subsection{Selective Attenuation Curve Variation with Inclination} \label{attenuation curve}
The selective attenuation is related to the total-to-selective attenuation, $k(\lambda)$, through the following relation
\begin{equation}\label{eq:k_def}
k(\lambda)= fQ(\lambda)+R_V \,,
\end{equation}
where $f$ is a constant required to make $k(B)-k(V)\equiv1$,
\begin{equation}\label{eq:f_def}
f= \frac{1}{Q_{\mathrm{eff}}(B)-Q_{\mathrm{eff}}(V)} \,,
\end{equation}
and $R_V$ is the total-to-selective attenuation in the $V$ band, which is the vertical offset of the curve from zero at 5500~\AA\ (i.e., $R_V\equiv k(V)$). We assume $B$ and $V$ bands to be 4400~\AA\ and 5500~\AA, respectively. The term $f$ is accounting for differences in the reddening between the ionized gas and the stellar continuum and can also be expressed as 
\begin{equation}\label{eq:f_def_ebv_ratio}
f = \frac{k(H\beta)-k(H\alpha)}{E(B-V)_{\mathrm{star}}/E(B-V)_{\mathrm{gas}}} \,,
\end{equation}
where $k(H\beta)$ and $k(H\alpha)$ are the values for the intrinsic extinction curve of the galaxy and \textit{not} from the attenuation curve. As such, the quantity $fQ(\lambda)$ represents the true wavelength-dependent behavior of the attenuation curve on the stellar continuum. For reference, the average attenuation curve of the entire $1.1<D_n4000<1.3$ UV-optical sample (i.e., without $r_{90}>4\arcsec$ criterion) has a value of $f=2.40\substack{+0.33 \\ -0.29}$ \citep{battisti16}.

The value of $f$ for the average selective attenuation curve of each subpopulation, based on the fits of the previous section, are outlined in Table~\ref{Tab:Q_vs_param} and \ref{Tab:Q_IR_vs_param}. The uncertainty shown reflects the maximum and minimum values from fits using individual $Q_{n,r}(\lambda)$. We find that the value of $f$ is lower for the decreasing axial ratios (higher inclinations), which from equation~(\ref{eq:f_def_ebv_ratio}) implies that \EBratio\ becomes slightly larger toward unity at higher inclination. This statement assumes the average intrinsic extinction curve among each bin is comparable, which is a reasonable assumption. If a MW extinction curve is assumed \citep[$k(\mathrm{H}\alpha)-k(\mathrm{H}\beta)=1.257$;][]{fitzpatrick99}, then \EBratio=0.47 and 0.55 for the highest and lowest axial ratio cases of the UV-optical sample, respectively. This indicates that there is a smaller difference between the reddening of the stellar continuum and ionized gas for more inclined galaxies. This effect is likely a result of the geometry, because both components will be affected more by the diffuse dust component along the line of sight in the edge-on scenario. We discuss this in more detail in Section~\ref{lit_compare}.

\begin{figure}
\begin{center}
\includegraphics[scale=0.5]{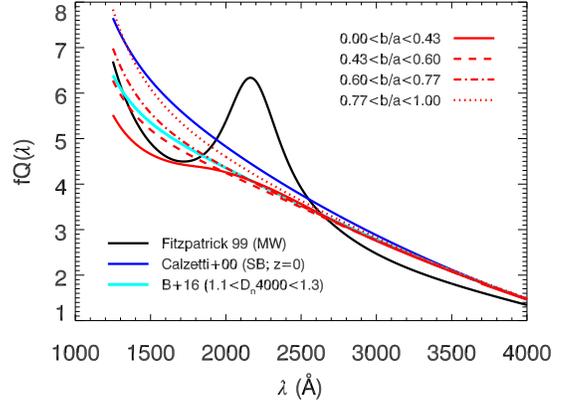}
\end{center}
\vspace{-0.2cm}
\caption{Normalized selective attenuation curve $fQ(\lambda)$ derived from our inclination subsets (red lines). The term $f$ is required to make the curve have $k(B)-k(V)\equiv1$. The cyan line is the average attenuation curve of the $1.1<D_n4000<1.3$ sample \citep[from which the $b/a$ subpopulations are drawn][]{battisti16}, the solid blue line is the starburst selective attenuation curve \citep{calzetti00}, and the solid black line is the MW extinction curve \citep{fitzpatrick99}. The shape of the curves in the UV is seen to become shallower at lower axial ratios (higher inclinations). The highest axial ratio subpopulation has a shape comparable to the starburst attenuation curve.
 \label{fig:fQ_compare}}
\end{figure}

We compare the selective attenuation curves of our UV-optical sample with this factor included in Figure~\ref{fig:fQ_compare}. For reference, we also show the average curve from \citet{battisti16} ($1.1<D_n4000<1.3$), the starburst attenuation curve of \citet{calzetti00}, and the MW extinction curve \citep{fitzpatrick99}. In this Figure, it can be seen that the selective attenuation curve becomes shallower in the UV with increasing inclination. Interestingly, for the most face-on subpopulation ($0.77<b/a<1.00$), the curve approaches a slope comparable to the starburst attenuation curve. We discuss possible reasons for these trends in Section~\ref{lit_compare}.

\begin{table*}
\begin{center}
\caption{Fit Parameters of $Q(\lambda)$ over $0.125~\micron\le\lambda<0.832~\micron$ as a Function of Axial Ratio \label{Tab:Q_vs_param}}
\begin{tabular}{ccccccc}
\hline\hline 
    range   & $f$ &  $p_0$    &  $p_1$    & $p_2$ &  $p_3$ & $E_b$  \\ \hline
$0.00<b/a<0.43$ & 2.28$\substack{+0.36 \\ -0.20}$ & -2.773 & 2.064 & -3.257$\times$10$^{-1}$ &  1.859$\times$10$^{-2}$ & 0.55 \\
$0.43<b/a<0.60$ & 2.38$\substack{+0.47 \\ -0.34}$ & -2.552 & 1.874 & -2.815$\times$10$^{-1}$ & 1.605$\times$10$^{-2}$ & \nodata \\
$0.60<b/a<0.77$ & 2.60$\substack{+0.56 \\ -0.42}$ & -2.333 & 1.710 & -2.585$\times$10$^{-1}$ & 1.540$\times$10$^{-2}$ & \nodata \\
$0.77<b/a<1.00$ & 2.70$\substack{+0.76 \\ -0.32}$ & -2.194 & 1.600 & -2.370$\times$10$^{-1}$ & 1.458$\times$10$^{-2}$ & \nodata \\ \hline 
\end{tabular}
\end{center}
\textbf{Notes.} The uncertainty in $f$ denotes the maximum and minimum values from fits using individual $Q_{n,r}(\lambda)$ for each subpopulation (see \S~\ref{attenuation curve}). The functional form of these fits are $Q_{\mathrm{fit}}(x) = p_0+p_1x+p_2x^2+p_3x^3$. These cases also have the constraint that $1.1<D_n4000<1.3$. The fit for the bump strength assumes $\lambda_0=2175$~\AA\ and $\Delta\lambda=470$~\AA\ \citep{fitzpatrick99}.
\end{table*}

\begin{table*}
\begin{center}
\caption{Fit Parameters of $Q(\lambda)$ over $0.63~\micron\le\lambda<2.1~\micron$ as a Function of Axial Ratio \label{Tab:Q_IR_vs_param}}
\begin{tabular}{cccccc}
\hline\hline 
  range   & $f$ &  $p_0$    &  $p_1$    & $p_2$ &  $R_V(2.85~\micron)$  \\ \hline 
$0.00<b/a<0.62$ & 2.21$\substack{+0.18 \\ -0.15}$ & -2.234 & 1.349 & -5.329$\times$10$^{-2}$ & 3.91$\substack{+0.65 \\ -0.77}$ \\ 
$0.62<b/a<1.00$ & 2.58$\substack{+0.16 \\ -0.14}$ & -1.766 & 1.020 & -1.827$\times$10$^{-2}$ & 3.64$\substack{+0.20 \\ -0.45}$ \\ \hline 
\end{tabular}
\end{center}
\textbf{Notes.} The uncertainty in $f$ and $R_V$ denote the maximum and minimum values from fits using individual $Q_{n,r}(\lambda)$ for each subpopulation (see \S~\ref{templates}). The functional form of these fits are $Q_{\mathrm{fit}}(x) = p_0+p_1x+p_2x^2$, where $x=1/\lambda$~($\micron^{-1}$). The normalization $R_V(2.85~\micron)$ is determined by extrapolating this function out to 2.85~$\micron$ and forcing $k(2.85~\micron)=0$ (see \S~\ref{normalization}). These subpopulations also have the constraint that $1.13<D_n4000<1.33$ and $9.1<\log[M_* (M_{\odot})]<9.9$.
\end{table*}

\subsection{Curve Normalization, \texorpdfstring{$R_V$}{RV}, and Comparison with Local Studies}\label{normalization}
We determine the normalization, $R_V$, for each of the UV-NIR subpopulations by extrapolating the NIR selective attenuation out to wavelengths where the total-to-selective attenuation is expected to approach zero, $k(\lambda\rightarrow\infty)\sim0$. This is often approximated as occuring at $\lambda\sim2.85$~\micron\ \citep[e.g.,][]{calzetti97b,gordon03,reddy15}. The values found for $R_V$ by extrapolating for the value of $fQ_{\mathrm{fit}}(\lambda)$ out to 2.85$~\micron$ and forcing $k(2.85~\micron)=0$ are shown in Table~\ref{Tab:Q_IR_vs_param}. We find that the lower axial ratio subpopulation has a slightly higher inferred $R_V$ value than the higher axial ratio, with values of $R_V=3.91\substack{+0.65 \\ -0.77}$ and 3.64$\substack{+0.20 \\ -0.45}$, respectively, where the uncertainty reflects the maximum and minimum values from fits using individual $Q_{n,r}(\lambda)$. However, this difference is not very significant given the large uncertainties. For reference, the average attenuation curve of the entire $1.13<D_n4000<1.33$ and $9.1<\log[M_* (M_{\odot})]<9.9$ UV-NIR sample (i.e., without $r_{90}>4\arcsec$ criterion) has a value of $R_V=3.67\substack{+0.44 \\ -0.35}$ \citep{battisti17}. Both of the values above are consistent with this value.

With the determination of $R_V$, it is possible for us to compare our curves to those of \citet{wild11}, which performed a similar analysis on the effects of inclination on the attenuation of \textit{GALEX}-SDSS-UKIRT matched local galaxies.  \citet{wild11} separate their sample of galaxies into two groups according to stellar mass surface density, $\mu_*=0.5M_{*,tot}/(\pi r_{50,z}^2)$, where $M_{*,tot}$ is the total stellar mass and $r_{50,z}$ is the Petrosian half-light radius in the z band. They adopt a break corresponding to the bimodal local galaxy population of \citet{kauffmann03b} that separates bulgeless ($\mu_*<3\times10^8~M_\odot~\mathrm{kpc}^{-2}$) and bulged ($\mu_*>3\times10^8~M_\odot~\mathrm{kpc}^{-2}$) galaxies. We find that $\sim$95\% of our parent sample would be classified into the lower stellar mass surface density (bulgeless) sample and therefore we will only compare to that sample. They sub-divide each group and examined the attenuation curves by different sSFR and the axial ratio ($b/a$). For clarity, we only reproduce the curves of different axial ratio and hold the sSFR fixed to their average value of $\log[\mathrm{sSFR~(yr}^{-1})]=-9.5$. 

\begin{figure*}
\begin{center}$
\begin{array}{cc}
\includegraphics[scale=0.5]{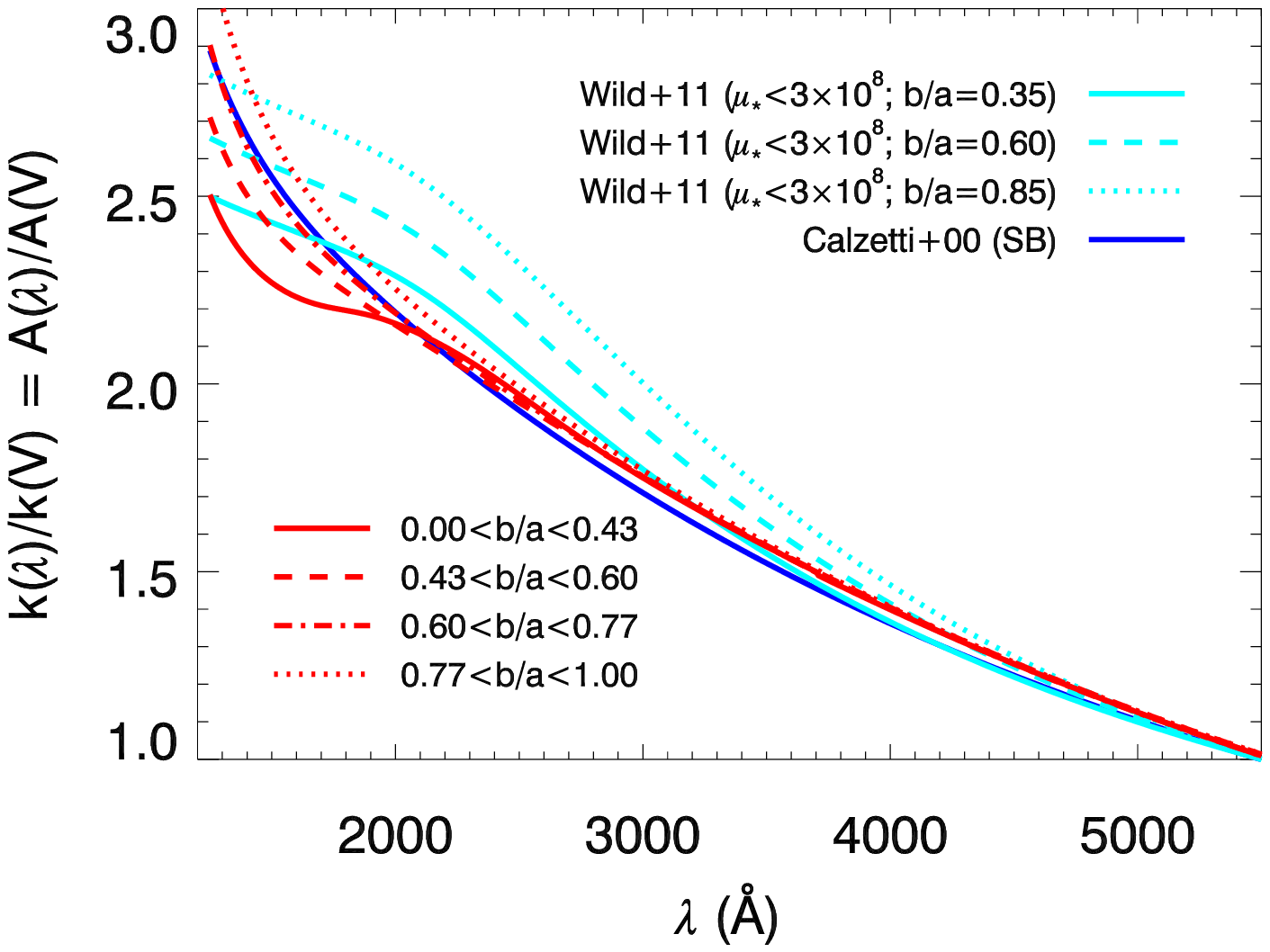} & \hspace{1cm}
\includegraphics[scale=0.5]{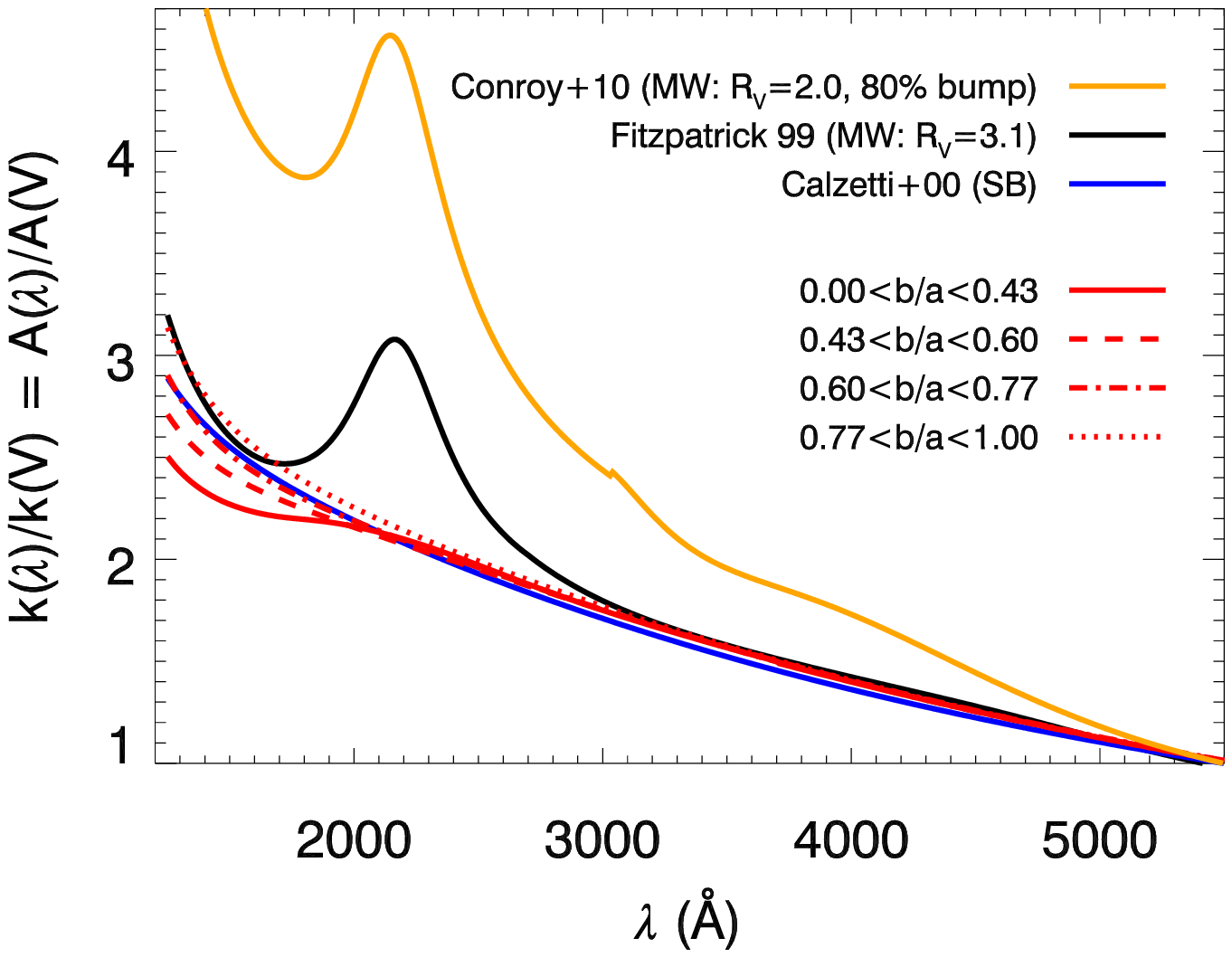} \\
\end{array}$
\end{center}
\vspace{-0.2cm}
\caption{\textit{Left:} The total-to-selective dust attenuation normalized at $V$-band, $k(\lambda)/k(V)$, for our axial ratio subpopulations (red lines) compared to those of \citet{wild11} for local galaxies (cyan lines). Only curves for the bulgeless ($\mu_*<3\times10^8~M_\odot~\mathrm{kpc}^{-2}$) galaxies of \citet{wild11} are shown, as the majority of our sample overlaps with this group. Both studies suggest a shallowing of the curves at higher inclinations, but the behavior of the 2175~\AA\ feature is different. These differences may be due to differences in source selection or methodology (see Section~\ref{normalization}). For reference, we also show the starburst selective attenuation curve \citep[solid blue line;][]{calzetti00}, which has $R_V=4.05$. \textit{Right:} $k(\lambda)/k(V)$ for our axial ratio subpopulations (red lines) compared to the preferred curve of \citet{conroy10} for local disk galaxies (orange line), which corresponds to a MW curve with an $R_V=2.0$ and a bump strength of 80\% that of the MW. The fiducial MW curve with $R_V=3.1$ is also shown for reference (black curve). \label{fig:wild_compare}}
\end{figure*}

We compare the total-to-selective attenuation curve normalized at $V$-band, $k(\lambda)/k(V)$\footnote{As a reminder, $k(\lambda)/k(V)=(fQ(\lambda)+R_V)/R_V$.}, of our UV-optical sample (Table~\ref{Tab:Q_vs_param}) to that of \citet{wild11} in Figure~\ref{fig:wild_compare} (\textit{Left}; cyan lines). It is worth noting that normalizing the curve in this way is equivalent to plotting the curves in the form of $A(\lambda)/A(V)$. For simplicity, here we adopt a single value $R_V=3.67$ for normalizing all of our UV-optical subsamples, although we find that adopting the slightly higher value of $R_V$ found above for the lower inclination subpopulations does not substantially alter the curves. Both studies find a shallowing in the slope at higher inclinations, but this extends to longer wavelengths in \citet{wild11}. Unlike our study, \citet{wild11} favor a 2175~\AA\ feature at all axial ratios with comparable strength, although the bump strength is seen to increase at higher inclinations in their bulged ($\mu_*>3\times10^8~M_\odot~\mathrm{kpc}^{-2}$) galaxies. Below we highlight differences between our study and \citet{wild11} that may account for the differing attenuation curves. 

First, this may be a consequence of the differences in the photometric aperture adopted in each study. \citet{wild11} use the entire galaxy for their photometry, which would lead them to enclose a larger amount of the diffuse dust disk relative our fixed aperture method (to remain close to the SDSS fiber). In our case, we tend to enclose a smaller fraction of the galaxy that is centered on the central star forming region. As will be discussed more in Section~\ref{lit_compare}, these two regions may have differing dust properties. Second, for their UV data \citet{wild11} utilize the \textit{GALEX} All-sky Imaging Survey (AIS) survey, as opposed to our use of the Medium Imaging Survey (MIS), which is shallower and may preferentially select bluer galaxies at low values of $\tau_B^l$ \cite{battisti16}. Thus, the galaxies selected in each study may be fundamentally different. Third, they adopt a pair matching method to determine their attenuation curves that is fundamentally different than our methodology and does not explicitly attempt to select for stellar population age. As discussed in \citet{battisti16} (and also \citet{wild11}), steeper apparent UV slopes for the attenuation curve can result from a mismatch in the mean stellar population ages of galaxies if dustier galaxies are on average intrinsically older, and therefore redder, than the less dusty galaxies. Such a trend is evident in our sample when comparing the values of $\tau_B^l$ and $D_n4000$ \citep[proxy for mean stellar age; see][]{battisti16}. Therefore, this could give rise to their higher inferred UV attenuation, although we note that this should be somewhat mitigated by their matching galaxies by sSFR and metallicity, both of which correlate with mean stellar population age. Both the second and third points made above can lead to steeper inferred attenuation curves in the UV region and we speculate these to be the primary causes of the differences seen in Figure~\ref{fig:wild_compare}.
 
We also show a comparison of $k(\lambda)/k(V)$ from our UV-optical sample to that of \citet{conroy10} in Figure~\ref{fig:wild_compare} (\textit{Right}; orange line). \citet{conroy10} performed a model-based derivation of the attenuation curve in \textit{GALEX}-SDSS matched local disk galaxies, selected to have S\'ersic index of $n\leq2.5$ and $9.5<\log(M/M_\odot)<10.0$. They compared the broadband photometry of galaxies of differing inclinations and inferred that they were best characterized using a MW curve with a $R_V=2.0$ and a bump feature with a strength of 80\% that of the MW. As can be seen in Figure~\ref{fig:wild_compare}, this curve is noticeably steeper than our derived curves, when all curves are normalized to the same value at
5500~\AA. As noted earlier, our sample is preferentially found to be disky, with 88\% of the parent sample having $n\leq2.5$, which makes the observed differences puzzling, but given the substantial differences in the adopted methodologies between these works it is difficult to determine sources of the discrepancy. However, we note that the study of \citet{conroy10} did not utilize NIR data and therefore their chosen value for $R_V$ (influencing the steepness of the curve when shown in this form) is not well constrained. Larger values of $R_V$ act to flatten the curve (i.e., closer to standard MW curve) and this might be the primary cause of the differences seen in Figure~\ref{fig:wild_compare}.

\section{Sources with GALEX spectroscopy}\label{galex_spectra}
Due to the limited UV sampling afforded through \textit{GALEX} photometry for analyzing the 2175~\AA\ feature, it is worthwhile to investigate the nature of this feature in an independent manner. To do this, we select galaxies in our sample that also have \textit{GALEX} spectroscopy. The \textit{GALEX} spectroscopic survey was originally planned to be comparable in scope to the imaging survey with three tiers of depth, but this was reduced to only the Medium Spectroscopic Survey (MSS) and Deep Spectroscopic Survey (DSS). The MSS and DSS cover $\sim8$ and $\sim2$ square degrees, respectively. For this section, we will consider sources within both the MIS (our parent sample) and the Deep Imaging Survey (DIS), where for the latter we use the Galaxy Multiwavelength Atlas from Combined Surveys \citep[GMACS;][]{johnson07a,johnson07b} dataset\footnote{\url{http://user.astro.columbia.edu/~bjohnson/GMACS/catalogs.html}}. We use the same selection criteria for the GMACS sample as the parent sample (i.e., UV $S/N>5$, emission lines $S/N>5$, SFG on BPT diagram, and $z\leq0.105$), which gives a sample of 476 galaxies. We locate sources with spectroscopy using the Mikulski Archive for Space Telescopes (MAST) CasJobs website\footnote{\url{http://galex.stsci.edu/casjobs/}}. More specifically, we select sources using the \texttt{isThereSpectrum} label in the \texttt{photoobjall} table and then obtain the reduced spectra from the \texttt{SpecFlux} table. The details and methodology of the spectral extraction techniques and the resulting grism data from the GALEX team is outlined in \citet{morrissey07}. GALEX performed spectroscopy in the FUV (1300-1820\AA) and the NUV (1820-3000\AA), with an average resolution ($\lambda/\Delta\lambda$) of about 200 and 118 ($\Delta\lambda\sim8$~\AA\ and 20~\AA), respectively. We note that the UV spectroscopy is an integrated (global) quantity in these galaxies, which is distinctly different from our previous aperture-based analysis.

\begin{figure}
\begin{center}
\includegraphics[scale=0.5]{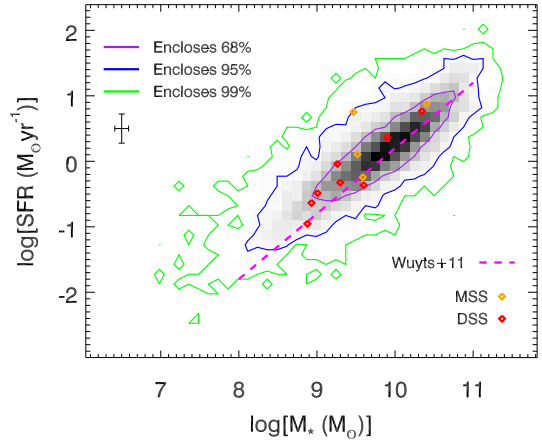}
\end{center}
\vspace{-0.2cm}
\caption{The location of GALEX spectroscopic sources relative to the galaxy main sequence. The sources appear to occupy a similar location on the galaxy main sequence as the parent sample (contour), implying that the sample should not be significantly biased. The average value of the main sequence for low redshift galaxies ($0.02<z<0.2$) from \citet{wuyts11} is shown as the dashed magenta line for reference. 
 \label{fig:MS_contour}}
\end{figure}

As a result of the limited coverage area of the MSS and DSS relative to MIS and DIS (0.8\% and 2.5\%, respectively), the overlap with the photometric sample is low. We identify 12 overlapping spectroscopic SFGs (4 in MSS and 8 in DSS) with good quality spectra with a median $S/N>5$ in both the FUV and NUV channels. We note that an additional MSS source was excluded as it was seen to be a close galaxy pair in the SDSS photometry. The average effective exposure time (\texttt{fuv\_effexp} and \texttt{nuv\_effexp}) for these galaxies is 13500~s and 40500~s for MSS and DSS, respectively. The location of these spectroscopic galaxies on the star-forming galaxy main sequence \citep[SFR vs. stellar mass; e.g.,][]{brinchmann04,wuyts11,cook14} is shown in Figure~\ref{fig:MS_contour}, where here we make use of the global values of SFR and stellar mass. The sources appear to occupy similar locations on the galaxy main sequence as the parent sample, implying that the sample should not be significantly biased, although they do tend to be a bit closer and larger in size.

The best discrimination on the 2175~\AA\ feature can be made on higher $\tau_B^l$ cases because $A_\lambda\propto E(B-V)_{\mathrm{star}}\propto\tau_B^l$ (assuming a fixed differential reddening factor). A concern with the analysis presented in this section is that $\tau_B^l$ is used to infer the global reddening for these galaxies, which is measured from the SDSS fiber and several these sources are much larger than the fiber size. We discuss an alternative to this approach at the end of this section.

\begin{figure*}
\begin{center} 
\includegraphics[scale=0.26]{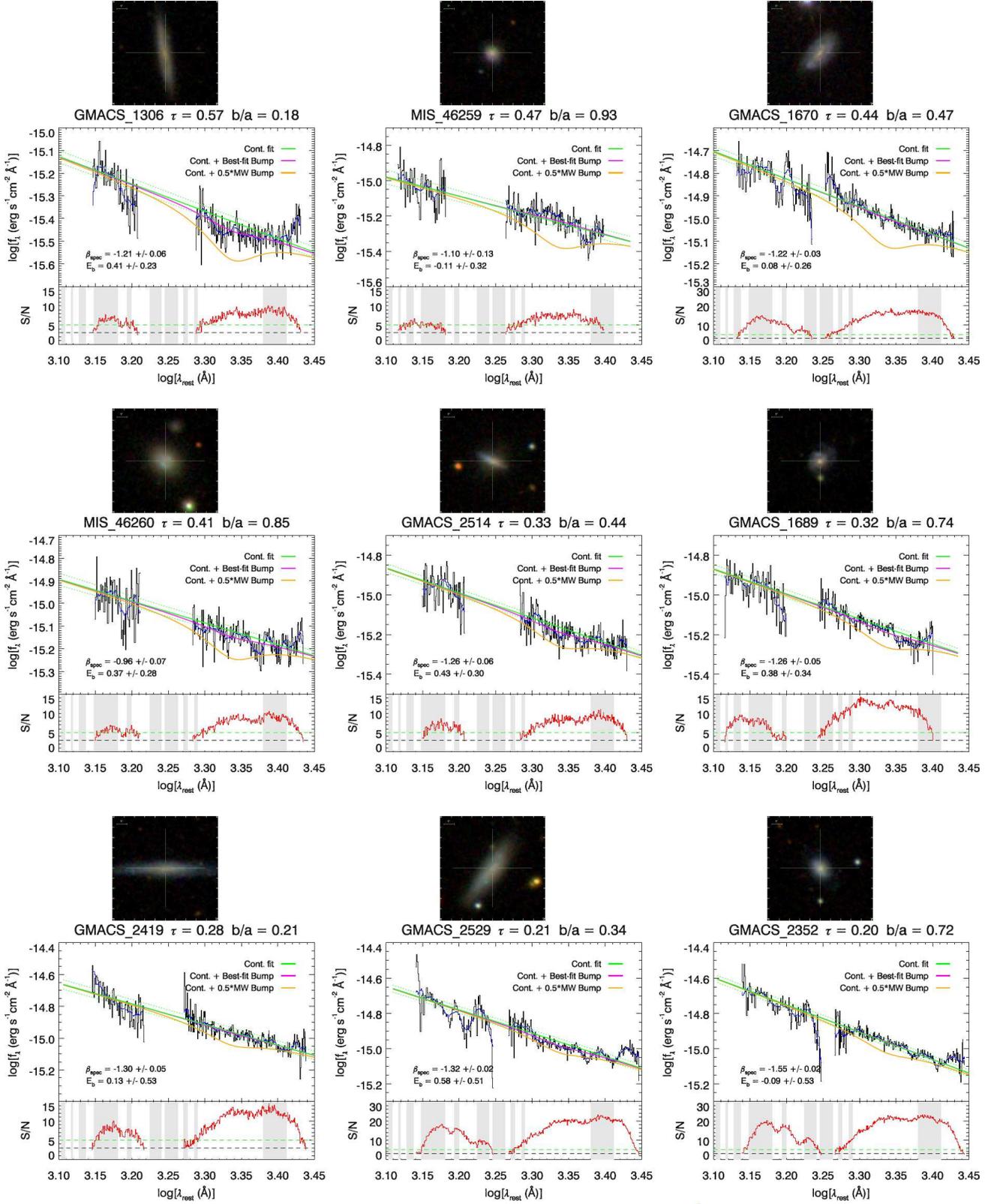}


\end{center}
\vspace{-0.4cm} 
\caption{SDSS thumbnails ($\sim$50$\arcsec$ $\times$ $\sim$50$\arcsec$) and GALEX spectra examples for the higher $\tau_B^l$ cases, where the best descrimination on the feature can be made ($A_\lambda\propto\tau_B^l$). The data is smoothed by 50~\AA\ (solid blue line) in order to fit the continuum. The smoothed data is fit using data with $S/N>5$ (dashed green line in lower panel) in the 10 windows presented in \cite{calzetti94} (gray regions in the lower panel). Using the continnum fit (solid green line) and uncertainty (dotted green line), we fit the  original spectrum (unsmoothed) for additional attenuation as the result of a 2175~\AA\ feature (solid magenta line). The best-fit UV continuum slope $\beta$ and bump strength are shown for each galaxy. For reference, we also show the expectation for a feature with a bump strength of half the MW value ($E_b=1.65$), and which is comparable to the LMC2 supershell. All cases are best-fit with features much lower than this value. 
\label{fig:GALEX_spectra}}
\end{figure*}

\begin{figure}
\begin{center}
\includegraphics[scale=0.5]{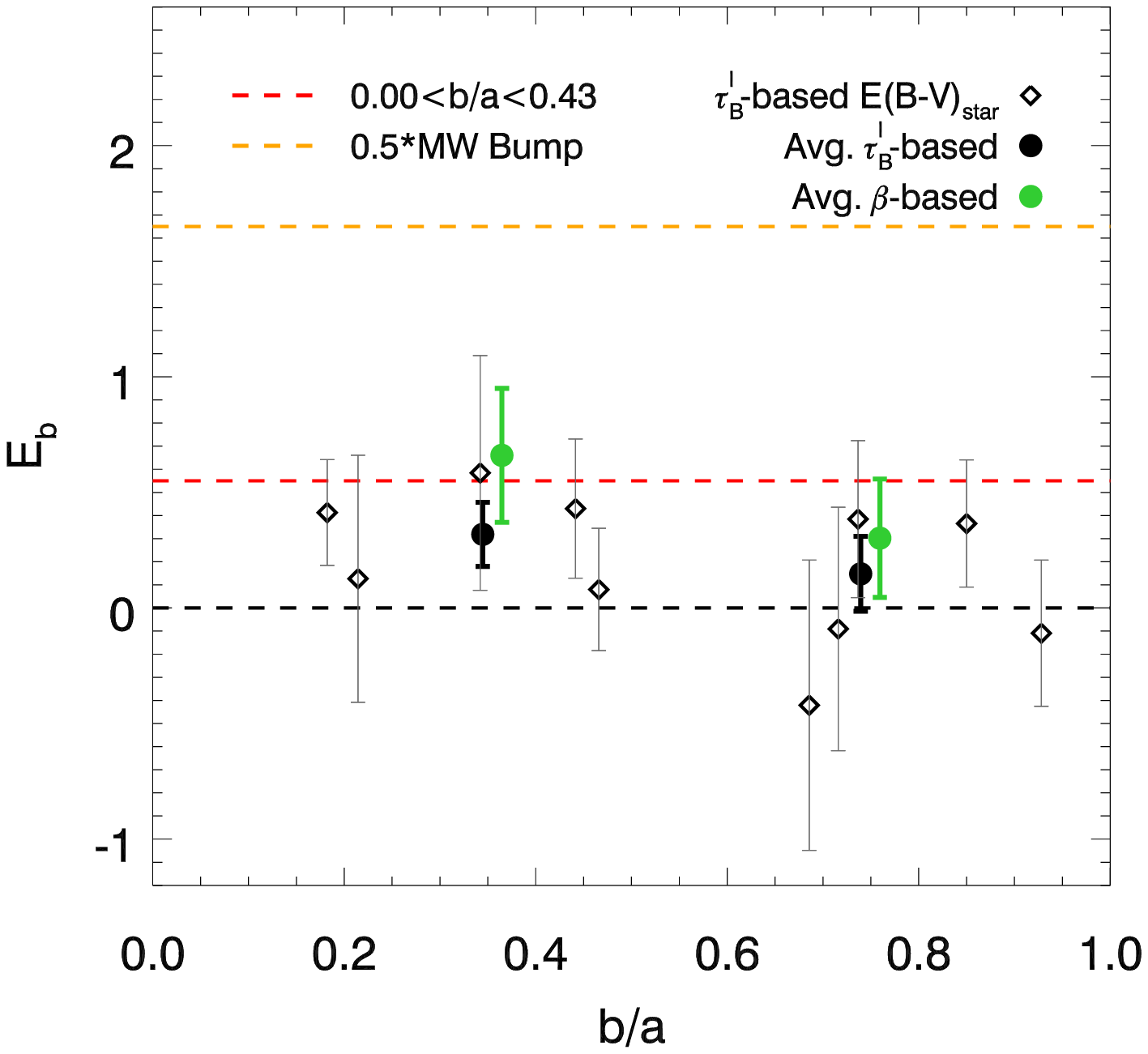}
\end{center}
\vspace{-0.2cm}
\caption{The best-fit bump strength $E_b$ as a function of axial ratio $b/a$ for our UV spectroscopic sources. Only sources with inferred $E(B-V)_{\mathrm{star}}>0.05$ based on $\tau_B^l$ are shown (10/12 cases) because lower values poorly constrain the bump strength and have very large error bars. The average values for sources above and below $b/a=0.5$ are also shown (filled black circles). There are no significant trends given the uncertainty, but a slight tendency for more inclined systems to have a larger bump value is present. For reference, The average values based on using $\beta$ to infer $E(B-V)_{\mathrm{star}}$ is also shown and leads to bump strengths that are larger by about 50\% from the $\tau_B^l$-based method (filled green circles; see Section~\ref{galex_spectra}) and is slightly offset in the x-axis for clarity. Regardless of the methodology, all galaxies show features much lower than the average MW value (orange dashed line shows 50\% of the MW value). The average bump value inferred from the NUV photometry for our highest inclination cases ($0.00<b/a<0.43$) is shown with the dashed red line. 
 \label{fig:b2a_vs_Eb}}
\end{figure}

The UV spectroscopy for the higher $\tau_B^l$ cases are shown in Figure~\ref{fig:GALEX_spectra}. For each case we fit the UV continuum, which is well represented as a power-law $F(\lambda)\propto\lambda^\beta$, using the 10 rest-frame wavelength windows defined in \citet{calzetti94}. These windows were used to measure the UV slope $\beta$ of starburst galaxies from spectra observed with the \textit{International Ultraviolet Explorer (IUE)} and are designed to avoid strong stellar absorption features, including the 2175~\AA\ feature. To get a better representation of the continuum, the data are smoothed by 50~\AA. We perform a linear least-squares fit (log-space) to the smoothed data with $S/N>5$ inside these windows using the MPFIT package \citep{markwardt09}. The uncertainty of this continuum is taken to be the formal fitting uncertainty and the dispersion of the smoothed data relative to the continuum added in quadrature. Using this continuum fit, we then determine the best-fit bump by taking a least-squares fit of the residuals (not smoothed) with attenuation from a 2175~\AA\ feature. This is determined using, 
\begin{equation}
F_{\mathrm{obs}}(\lambda) = F_{\mathrm{int}}(\lambda) 10^{-0.4E(B-V)_{\mathrm{star}}D(\lambda)} \,,
\end{equation}
where the color excess of the stellar continuum is determined from the Balmer optical depth,
\begin{equation}\label{eq:EBV_star}
E(B-V)_{\mathrm{star}} = \frac{1.086 \tau_B^l}{f} \,,
\end{equation}
and we use $f=2.40$ corresponding to the average value for SFGs from \citet{battisti16}. We fix the central wavelength and width of the bump feature to that of the MW ($\lambda_0=2175$~\AA\ and $\Delta\lambda=470$~\AA), leaving $E_b$ as the only free parameter. The resulting fits and $E_b$ values are illustrated in Figure~\ref{fig:GALEX_spectra}, where the uncertainty in $E_b$ is the formal fitting uncertainty and the continuum uncertainty added in quadrature. For reference, we also demonstrate the effect of using a fixed $E_b=1.65$ (50\% of the MW value, 99\% of the LMC2 supershell value). A feature of this strength can be ruled out for all cases. The bump strength as a function of the inclination is shown in Figure~\ref{fig:b2a_vs_Eb}. 

As mentioned above, the global $E(B-V)_{\mathrm{star}}$ may be smaller than that inferred from the SDSS fiber. Several studies have suggested that dust attenuation decreases with increasing radius \citep[e.g.,][]{munoz-mateos09,iglesias-paramo13,nelson16a}. If this is the case, then the average bump strength would increase slightly from the values presented above. As an independent estimate, we also use the UV slope as an indicator of the stellar reddening adopting the attenuation relationship in \citet[][equation~17]{battisti17}. This relations assumes an intrinsic UV slope of $\beta_0=-1.61$. This leads to bump strengths that are larger by about 50\% (demonstrated in Figure~\ref{fig:b2a_vs_Eb}). If instead, we adopted the attenuation relation for starburst galaxies \citep[][combining their equation~9 and 10]{calzetti00}, the resulting $E(B-V)_{\mathrm{star}}$ values are higher because this relation assumes $\beta_0=-2.1$ (providing smaller difference from the $\tau_B^l$-based method). Regardless of the adopted methodology, the statements made earlier that the bump strengths are significantly below the MW and LMC2 supershell values remain valid.

\section{Variation in \texorpdfstring{$\beta$}{beta}-\texorpdfstring{$\tau_B^l$}{tau} with Inclination}\label{atten_vs_inclin}
The variation of the UV slope $\beta$ with inclination also has important implications for the use of this quantity as an indicator of the total dust attenuation when other metrics are unavailable. Any additional trends with the 2175~\AA\ feature would further complicate the use of the region of $1950<\lambda<2300$~\AA\ in calculating $\beta$. A common method of dust corrections is to use of the tight positive correlation between $\beta$ and the ratio of IR to UV luminosity, termed the ``infrared excess'' ($IRX=L_{\rm{IR}}/L_{\rm{UV}}$), which is a proxy for the total dust attenuation, observed for starburst galaxies \citep{meurer99}. However, the $IRX-\beta$ correlation is seen to break down as one moves from starburst galaxies to more ``normal'' SFGs \citep{kong04, buat05, hao11}. This breakdown has been attributed to effects of evolved stellar populations, different star formation histories, variations in the dust/star geometry, and effects from the 2175~\AA\ feature \citep[e.g.,][]{boquien09, grasha13, popping17}, all of which impact $IRX$ and/or $\beta$ values and introduce scatter.

\begin{figure*}
\begin{center}
\includegraphics[scale=0.4]{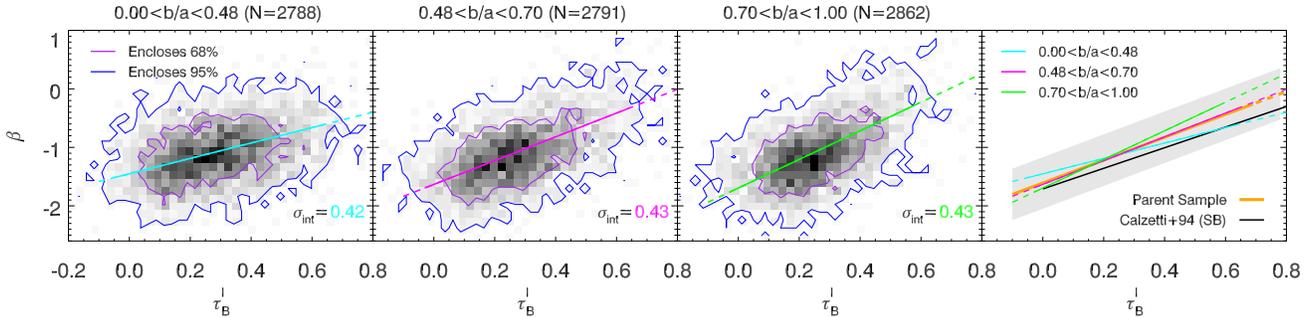}
\end{center}
\vspace{-0.2cm}
\caption{The $\beta$-$\tau_B^l$ relation for subpopulations of galaxies with different axial ratios, where $\beta$ is the UV power-law index after correcting for stellar absorption features \citep{battisti16}. The slope and offset of the relationship varies noticeably with the inclination, with flatter slopes and redder zero-points at lower axial ratios (higher inclination). The right panel shows the comparison to the parent sample and the local starburst relation \citep{calzetti94}, where the gray filled region denotes the intrinsic dispersion of the parent sample ($\sigma_{int} = 0.44$).
\label{fig:beta_tau_subsample}}
\end{figure*}

Following a similar analysis as presented in \citet{battisti16}, we examine the variation in the $\beta$-$\tau_B^l$ relation for galaxies with inclination to identify its behavior and determine if it is responsible for any of the intrinsic scatter observed. Here we will use the corrected version of $\beta_{\rm{GLX}}$ UV slope from \citet[][see Figure~12]{battisti16}, which accounts for the absorption features influencing the wide \textit{GALEX} filters. This essentially acts to shift the values of $\beta_{\rm{GLX}}$ bluer (more negative) by -0.15. The results are shown in Figure~\ref{fig:beta_tau_subsample}. We chose the subpopulations to consist of roughly one-third of the sample (each $\sim2800$ galaxies). We note regions with low sampling of galaxies with dashed lines. Looking at Figure~\ref{fig:beta_tau_subsample}, it can be seen that there are notable differences with the axial ratio, although interestingly the intrinsic scatter is not reduced relative to the entire parent sample ($\sigma=0.44$). The slope and offset of the relationship varies with the inclination, with flatter slopes and redder zero-points at lower axial ratios (higher inclination):
\begin{equation}
\begin{array}{lr}
\beta = (1.32\pm0.05) \tau_B^l -(1.45\pm0.02) & 0.00<b/a<0.48 , \\
\beta = (2.03\pm0.05) \tau_B^l -(1.63\pm0.02) & 0.48<b/a<0.70 , \\
\beta = (2.44\pm0.06) \tau_B^l -(1.69\pm0.02) & 0.70<b/a<1.00 .
\end{array}
\end{equation}

The trends observed in Figure~\ref{fig:beta_tau_subsample} can be attributed to the shallowing of the UV slope for lower axial ratios and also the addition of the weak bump feature (affecting our NUV data). Both of these effects will act to keep the UV slope bluer at higher attenuation values ($\tau_B^l$). It is less clear what is driving the differences in the offset (zero-point) among the relationships, although this is not particularly significant given the large intrinsic scatter. Another factor that may influence this is the change in differential reddening between the ionized gas and stellar continuum of the galaxies as a function of inclination. For example, at a fixed value of $\tau_B^l$ (fixed ionized gas attenuation), the higher inclination galaxies (with lower value of $f$) experience $E(B-V)_{\mathrm{star}}$ values that are 20\% larger than the lower inclination galaxies.

\section{Discussion}
\subsection{Our Results in the Broader Context of Previous Studies}\label{lit_compare}
In this section, we outline possible explanations for the observed trends with inclination, as well as place them in the context of results found in previous studies. Several studies have proposed that the dust content of galaxies can be well described by two components \citep[e.g.,][]{calzetti94, charlot&fall00, daCunha08, wild11}; one associated with short-lived dense clouds where massive stars form HII regions giving rise to the ionized gas emission (with a low covering fraction) and another associated with the diffuse interstellar medium (with a high covering fraction) that provides the attenuation on the stellar continuum. The two component model was originally developed to account for the fact that the color excess of the stellar continuum $E(B-V)_{\mathrm{star}}$ tends to be lower than that of the ionized gas emission $E(B-V)_{\mathrm{gas}}$ by roughly a factor of two \citep[e.g.,][]{calzetti00, kreckel13}, and has been extended to explain other trends as well \citep{wild11}. We argue below that this picture can also naturally explain many of the trends that are seen in this study.

The first trend that we consider is the attenuation curves becoming shallower in the UV with increasing inclination. This trend, also observed in \citet{wild11}, is expected to occur if the effective optical depth increases with inclination, leading to grayer overall attenuation \citep[e.g.,][]{pierini04, inoue05, chevallard13}. The fact that only the UV region appears to experience substantial changes with inclination is interesting. The UV emission is dominated by young stellar populations, which are likely to be located at small disk scale-lengths and heights and reside inside the diffuse dust component \citep[however, continuum emission from stars with ages $t<10$~Myr, which provide the line emission, will be linked to the dense cloud component;][]{charlot&fall00}. Therefore, we can expect that the effective optical depth toward young stars will depend noticeably on inclination, experiencing a minimum when viewed face-on and a maximum when viewed edge-on. In contrast, we expect the emission at optical and NIR wavelengths to be dominated by intermediate and old stellar populations, respectively, and these will be more homogeneously distributed across the disk, with older stars tending to reside at higher disk scale heights. Since the dust scale-height is typically found to be smaller than that of the stellar component, \citep[e.g.,][]{xilouris99, bianchi07, deGeyter14}, a fraction of these older stellar populations can reside outside of the diffuse dust component and would thus be less affected by changes in inclination. Therefore, as long as the effective optical depth does not become very large, the shape of the attenuation curve at optical and NIR wavelengths could remain unchanged (although the total attenuation still increases at higher inclinations). It can be seen in Figure~\ref{fig:tau_vs_b2a} that the $\sim$68\% and 95\% upper boundary of our sample is $\tau_B^l=0.4$ and 0.6, respectively. Assuming a differential reddening factor of $f=2.40$ and the average attenuation curve in \citet{battisti17}, these values  correspond to $A_V=0.66$ and 1.00, respectively. Therefore, the inferred optical depth for the majority of our sample is $\tau_V<1$ ($\tau(\lambda)=0.921A(\lambda)$), indicating that this assumption may be reasonable.

The second aspect we examine is tentative trend of an increase in the value of $R_V$ for the higher inclination UV-NIR galaxy subpopulation. This increase could be influenced by differences in the average grain size with the line of sight as there will be a higher probability of intersecting dense cloud components (low covering fraction) in an edge-on scenario. Extinction measurements in the MW toward denser environments (clouds) tend to have larger values of $R_V$ and shallower slopes and is likely associated with grain growth \citet{draine03}. This effect may also be linked to the flattening of the UV shape with inclination discussed in the previous paragraph. Utilizing larger samples in the future, for finer $b/a$ binning, will provide a more robust characterization of this effect.

The third trend to account for is the reddening between the ionized gas and stellar continuum becoming more similar (\EBratio\ becoming slightly larger toward unity) at higher inclination, which is also observed in \citet{wild11}. As mentioned above, the ionized gas is observed to experience a larger inherent reddening (optical depth) than the stellar continuum. In the two component model, we expect the difference in reddening to appear most pronounced in a face-on scenario, because the stellar continuum will experience the lowest optical depth from the diffuse component (and the ionized gas attenuation is relatively insensitive to inclination; see Section~\ref{method_attenuation}). At larger inclinations, the total optical depth of the diffuse component will increase for both ionized gas and stellar components but this has a much larger impact on the reddening of the continuum. Therefore, one expects the level of reddening between the two components to become closer at the highest inclinations. 

Next we consider the excess NUV attenuation in our most inclined subpopulation, which we consider to be the result of 2175~\AA\ feature with $\sim$20\% the strength of the MW. Given our crude wavelength sampling in the UV to determine the baseline surrounding the 2175~\AA, it is unclear if this feature is entirely absent in the other subpopulations but we expect that it is weaker given the steepening of the FUV for more face-on galaxies (higher baseline). Additionally, an independent examination of a handful of our sources in UV spectroscopy also suggests the 2175~\AA\ feature in attenuation curves are weaker than both the MW and LMC2 supershell extinction curves. Interestingly, for our lowest inclination sample the shape of the attenuation curve approaches that of the starburst attenuation curve, which is definitively seen to lack the 2175~\AA\ feature \cite{calzetti94}. Theoretical studies utilizing dust radiative transfer have mainly focused on adopting a single intrinsic extinction curve for their dust. When doing so, they predict that the 2175~\AA\ bump strength should decrease with increasing inclination owing to the increase in the effective optical depth leading to grayer overall attenuation \citep[e.g.,][]{pierini04, inoue05}. \citet{chevallard13} considered the effects of a two component dust scenario, but this focused only on attenuation at optical-NIR wavelengths. A key ingredient missing from previous radiative transfer models is in the role that higher intensity UV fields may have on the carrier of the 2175~\AA\ feature. This has been postulated as the primary reason the starburst attenuation curve lacks a bump feature \cite{gordon97,fischera&dopita11}, and is also evident in the extinction curves toward higher star-forming regions \citep[e.g., LMC2 supershell/30 Dor][]{gordon03}. If the dust surrounding star-forming regions is processed in a manner that destroys the carrier of the 2175~\AA\ feature, this could explain the lack of the feature in our more face-on galaxies, as the primary dust attenuation on these regions will be due to dust in its immediate vicinity (due to its small scale-height). In contrast, we expect that if the 2175~\AA\ feature carrier is present in the more general diffuse dust component that is outside the influence of the star-forming region, then the bump would be most prevalent in the UV continuum when galaxies are viewed nearly edge-on. Such findings are qualitatively consistent with \citet{wild11} and also \citet{kriek&conroy13} (although the latter study also suggests that edge-on galaxies preferentially have steeper slopes, which conflicts with the findings of this study). We also highlight again that our use of fixed-apertures located on the central star-forming regions (SDSS fiber) may lead us to preferentially examine regions experiencing higher dust processing than is typical in the outer regions of the disk. For galaxies in our study at high inclinations, a larger fraction of the outer region diffuse dust is likely included within these apertures and this may be an important factor in our curve behavior (this may also account for the bump being present at all inclinations in \citep{wild11}, see Figure~\ref{fig:wild_compare}, because they use total photometry).

Finally, we also remark on the prevalence of the bump feature in observations. Due to the methodology used in this study of averaging galaxy SEDs to quantify average attenuation curves, we are unable to constrain what fraction of our galaxy sample contains significant 2175~\AA\ features. Doing so would require performing SED fitting for individual galaxies in the sample, a task that would be highly model dependent, and is beyond the scope of this paper. However, if it is the case that this feature appears to be preferentially stronger in systems that are viewed at high inclinations, then one might expect the strongest instances of the bump to occur in $\sim$25\% of cases. This could account for a similar fraction of galaxies with significant bump strengths ($\gtrsim$30\% MW) being measured by \citet{noll09a, buat12, kriek&conroy13} and others. Unfortunately, the role of inclination with bump strength of individual galaxies has not been examined in much detail to date. Performing such studies in the future could greatly improve our ability to provide more appropriate treatment of the 2175~\AA\ feature when performing attenuation corrections.  

\section{Conclusions}
Using a sample of $\sim$10,000 local SFGs, we quantified the influence that inclination has on the shape and normalization on average dust attenuation curves of the central star forming regions. Below, we summarize our main conclusions:

\begin{itemize}
\item Attenuation curves of the central active regions of SFGs become shallower at UV wavelengths with increasing inclination, whereas the shape at longer wavelengths remains unchanged.
\item The amount of differential reddening between the stellar continuum and ionized gas varies with inclination, with \EBratio\ becoming slightly larger toward unity at higher inclination (i.e., smaller differences in reddening). 
\item The most edge-on subpopulation ($b/a<0.42$) exhibits a NUV excess in its average selective attenuation. If interpreted as a 2175~\AA\ feature, this is best fit by a feature with a bump strength of 17-26\% of the MW value (34-52\% of the LMC2 supershell value) depending on assumptions of the feature central wavelength and FWHM. No excess is apparent in the average attenuation curves of our lower inclination subpopulations ($b/a>0.42$).
\item A small difference is observed in the value of $R_V$ with inclination, such that the higher inclination subpopulation has a slightly higher $R_V$, although this is poorly constrained given the more limited sample with NIR data.
\item An independent examination of a small number of UV spectroscopic sources in our sample supports our previous finding that the 2175~\AA\ feature in attenuation curves are weaker than both the MW and LMC2 supershell extinction curves.
\item The inclination has a noticeable impact on the $\beta$-$\tau_B^l$ relation due to the change in the UV attenuation profile. This may add an additional complication to the use of the UV slope for the purpose of making attenuation corrections (e.g., through the IRX-$\beta$ relation).
\end{itemize}

Overall, the trends observed are consistent within the common picture that dust in galaxies can be separated into two dust components, one associated with very young stars with a low covering fraction (dense clouds) and another associated with the diffuse interstellar medium with a large covering fraction \citep{charlot&fall00}. These results are, for the most part, in qualitative agreement with an analogous study by \citet{wild11}. However, an inclination dependence of the 2175~\AA\ feature may additionally require that the carriers are preferentially destroyed regions of high star-forming activity. We caution that the results presented in this work are determined by averaging the SEDs of a large number of galaxies, and that variation in the dust properties of individual galaxies is likely to result in large case-by-case variation. Performing more in-depth analysis of individual galactic systems, with greater UV coverage than that afforded with \textit{GALEX} photometry, will be necessary to improve upon our understanding of the factors influencing this scatter. 

\section*{Acknowledgments} The authors thank the anonymous referee whose suggestions helped to clarify and improve the content of this work. AJB also thanks K. Grasha for comments that improved the clarity of this paper and T. M. Tripp for suggestions that led to the analysis of \textit{GALEX} spectroscopic sources.

Part of this work has been supported by the National Aeronautics and Space Administration (NASA), via the Jet Propulsion Laboratory Euclid Project Office, as part of the ``Science Investigations as Members of the Euclid Consortium and Euclid Science Team'' program.

This work is based on observations made with the NASA Galaxy Evolution Explorer. \textit{GALEX} is operated for NASA by the California Institute of Technology under NASA contract NAS5-98034.
This work has made use of SDSS data. Funding for the SDSS and SDSS-II has been provided by the Alfred P. Sloan Foundation, the Participating Institutions, the National Science Foundation, the US Department of Energy, the National Aeronautics and Space Administration, the Japanese Monbukagakusho, the Max Planck Society and the Higher Education Funding Council for England. The SDSS website is http://www.sdss.org/. The SDSS is managed by the Astrophysical Research Consortium for the Participating Institutions.
This work has made use of data obtained with the United Kingdom Infrared Telescope. UKIRT is supported by NASA and operated under an agreement among the University of Hawaii, the University of Arizona, and Lockheed Martin Advanced Technology Center; operations are enabled through the cooperation of the East Asian Observatory. When the data reported here were acquired, UKIRT was operated by the Joint Astronomy Centre on behalf of the Science and Technology Facilities Council of the U.K.
This publication makes use of data products from the Two Micron All Sky Survey, which is a joint project of the University of Massachusetts and the Infrared Processing and Analysis Center/California Institute of Technology, funded by the NASA and the National Science Foundation.

\bibliography{AJB_bib}

\end{document}